\documentclass[journal]{IEEEtran}

\IEEEoverridecommandlockouts 

\usepackage{cite}

\ifCLASSINFOpdf
   \usepackage[pdftex]{graphicx}
   \graphicspath{{./Images/}}
   \DeclareGraphicsExtensions{.pdf,.jpeg,.png}
\else
\fi
\usepackage[cmex10]{amsmath}
\usepackage{hyphenat}
\usepackage[cmex10]{amsmath}
\usepackage{amssymb}
\usepackage{verbatim}
\usepackage{amsfonts}
\usepackage{subfig}
\usepackage[update,prepend]{epstopdf}
\usepackage{float}
\usepackage[usenames,table,dvipsnames]{xcolor}
\usepackage[hidelinks]{hyperref}
\usepackage{tabularx}
\usepackage{bm}
\usepackage{siunitx}
\usepackage{enumerate}
\usepackage{multirow}
\usepackage{booktabs,colortbl}
\usepackage{algorithm,algpseudocode}
\usepackage{setspace}
\usepackage{diagbox}
\usepackage{mathrsfs}
\usepackage[export]{adjustbox}

\newcommand{\matr}[1]{\mathbf{#1}}
\newcommand{\vect}[1]{\bm{#1}}

\allowdisplaybreaks

\begin{document}

\title{Controlled Power System Separation Using Generator PMU Data and System Kinetic Energy}
\author{Ilya~Tyuryukanov,~\IEEEmembership{Member,~IEEE},          
        Jorrit A. Bos,       
        Mart A. M. M.~van~der~Meijden,~\IEEEmembership{Senior Member,~IEEE,}
        Vladimir~Terzija,~\IEEEmembership{Fellow,~IEEE}, 
        and~Marjan~Popov,~\IEEEmembership{Fellow,~IEEE} 
\thanks{Manuscript submitted on 1 July 2022. This research work was financially supported by the Dutch Scientific Council NWO in collaboration with TSO TenneT, DSOs Alliander and Stedin, VSL, and General Electric in the framework of the Energy System Integration \& Big Data program under the project ReSident, no. 647.003.004.}%
\thanks{I. Tyuryukanov and M. Popov are with Delft University of Technology, Delft 2628CD, Netherlands. Since October 1, 2021, I. Tyuryukanov is also with Siemens Energy AG, Freyeslebenstr. 1, Erlangen 91058, Germany (e-mail: ilya.tyuryukanov@ieee.org, m.popov@tudelft.nl).}%
\thanks{J. Bos and M.\,A.\,M.\,M.~van~der~Meijden are with TenneT TSO B.V., Utrechtseweg 310, Arnhem 6812AR, Netherlands (e-mail: \{Jorrit.Bos, mart.vander.meijden\}@tennet.eu). M.\,A.\,M.\,M.~van~der~Meijden is also with Delft University of Technology, Delft 2628CD, Netherlands.}%
\thanks{V. Terzija is with Key Laboratory of Power System Intelligent Dispatch and Control of Ministry of Education, School of Electrical Engineering, Shandong University, Jinan 250061, China (email: terzija@ieee.org)}}%


\markboth{Journal of \LaTeX\ Class Files,~Vol.~14, No.~8, August~2015}%
{Shell \MakeLowercase{\textit{et al.}}: Bare Demo of IEEEtran.cls for IEEE Journals}
\maketitle

\begin{abstract}
With the growing number of severe system disturbances and blackouts around the world, controlled system separation is becoming an increasingly important system integrity protection scheme (SIPS) to save the electric power system from a complete or partial disintegration. A successful controlled splitting approach should at least tackle the following two well-known and interrelated problems: "when to split?" and "where to split?". Multiple previous publications consider these problems separately, and even those pursuing a combined approach propose solutions with limited applicability. In this paper, we are proposing a novel PMU-based detector of loss of synchronism that utilizes generator PMU data to promptly detect rotor angle instabilities over a wide area. Moreover, we are showing how our loss of synchronism detection principle can be coupled with the known controlled splitting techniques to form an integrated defense scheme against unintentional loss of synchronism. The performance of this wide-area SIPS is demonstrated on the IEEE 39-bus test power system for various types of unstable conditions. 
\end{abstract}

\IEEEpeerreviewmaketitle

\section{Introduction}
\label{sec:intro}
In recent years, the increasing low-inertia intermittent power generation and the growing frequency of extreme weather events have contributed to multiple notable blackouts and system disturbances around the globe \cite{Kundu.2016}. As these trends are expected to continue and to overlap with the preexisting challenges to secure grid operation such as electricity deregulation, the demand for novel adaptive power system emergency control and protection solutions continues to grow as well. 

Controlled system separation is an important SIPS \cite{Madani.2010} aiming to prevent system collapse due to wide-area instability by separating the system into a set of non-interacting islands. A typical situation requiring such a control action is loss of synchronism between some system parts, which is also known as wide-area out-of-step (OOS) condition. Some recent notable examples of such a condition are: the two instances of system separation in the synchronous power grid of Continental Europe that occurred in 2021 \cite{ENTSOE.2021a,ENTSOE.2021b}, the Brazilian blackout of 2018 \cite{Alhelou.2019}, and the South Australian blackout of 2016 \cite{Alhelou.2019}. Currently, such events are normally handled by OOS relays that disconnect the protected element (e.g., a transmission line or generator) once an unstable power swing is recognized from the local measurements. However, such relays are difficult to tune with respect to varying system conditions \cite{Ariff.2016}, and their cumulative OOS tripping response may lead to an excessive load and generator shedding. 

Unlike traditional OOS relays, controlled power system separation should  coordinate the OOS detection and the subsequent line tripping actions. In the controlled splitting literature \cite{Sun.2011,Trodden.2014,Ding.2018}, the OOS detection is known as the "when to split" problem, and the splitting cutset selection is known as the "where to split" problem \cite{Sun.2011,Kundu.2016}. In most of studies, these problems are considered independently from each other. 

The \emph{where} problem typically aims to isolate the diverging generator groups while minimizing load and generator shedding within the formed network islands and enhancing their stability. This results in a mixed-integer nonlinear programming (MINLP) problem that is often approximated by mixed-integer linear programming (MILP)  \cite{Trodden.2014}, constrained graph partitioning \cite{Tyuryukanov.2018b}, or ordered binary decision diagrams (OBDDs) \cite{Zhao.2003} due to its high computational complexity.

The \emph{when} problem typically aims to detect OOS and to determine which generator groups should be separated from each other for the given instability. The common practical approach to this problem involves declaring instability once the angle difference between a pair of power system areas reaches a certain threshold and subsequently separating these areas \cite{Ohura.1990,Sauhats.2017}. However, it is acknowledged that the unstable angle differences between the areas are hard to select as they often differ for various disturbances \cite{Ohura.1990}. Other approaches from the literature include artificial intelligence (AI) based techniques such as decision trees \cite{Senroy.2006} or artificial neural networks (ANN) \cite{Paul.2020}, as well as various methods based on equal area criterion (EAC) \cite{Paudyal.2010,Alinezhad.2017,Ayer.2020,Tealane.2022} and, more broadly, on direct stability methods \cite{Ding.2018}. The drawbacks of AI-based techniques include the lack of tractability in many AI algorithms, the dependence on large amounts of training data, and the lack of robustness against new system conditions that did not appear in the training set. The applicability of EAC-based methods is limited to situations close enough to the evaluation of \emph{first swing stability} of a single machine infinite bus system (SMIB), as EAC is a particular case of direct stability methods \cite{Pai.1989,Chiang.2011,Padiyar.2013,Rimorov.2018} that is generally invalid for multimachine power systems equipped with regulators. Meanwhile, the use of more general direct stability methods is further discussed in Section \ref{sec:trastess}.

This paper is primarily focused on the \emph{when} problem by proposing a novel wide-area OOS protection scheme. Thus, it is very loosely related to our previous results presented in \cite{Tyuryukanov.2018b,naglic.2019,Tyuryukanov.2020} that are mostly focused on the \emph{where to split} problem and generator coherency. However, we also propose a clear and practical coupling between the \emph{when} and \emph{where} problems, thus outlining a complete PMU-based controlled splitting scheme. Our scheme utilizes time-synchronized generator angles and frequencies to compute several stability indicators that are used together for the wide-area OOS detection. Here the PMU time synchronization is extremely crucial, because without it various wide-area signals cannot be combined inside equations evaluated in real-time. The individual contributions and main results of the paper can be summarized as follows:
\begin{enumerate}
\item A novel wide-area OOS detector based on system kinetic energy and diverging generator angles (Section \ref{sec:WkeGMM}). 
\item A novel wide-area OOS detector based on generator phase portraits and wide-area stability indices, which mutually support each other (Section \ref{sec:PP}). 
\item A novel wide-area OOS detector that is specialized on undamped power oscillations (Section \ref{sec:Wke}). 
\item A novel real-time algorithm to identify which groups of generators are moving apart once OOS has been detected (Section \ref{sec:sss}). It avoids generator coherency identification that requires long observation windows and sophisticated algorithms for clustering complex time series.
\item A test framework to evaluate the OOS detection speed of the proposed method (Section \ref{sec:casestud}).
\end{enumerate}

Unlike detectors based on direct stability methods \cite{Ding.2018} or EAC \cite{Alinezhad.2017}, our OOS detection method is inherently signal based and requires no real-time information about the network model. Our scheme is equally suitable for OOS due to short circuit faults and undamped power swings, which is favorable compared to the methods that rely on modal information \cite{Sun.2011} or long observations \cite{Salimian.2018}. Additionally, our scheme can handle both a single generator OOS \cite{Alinezhad.2017} and the loss of synchronism between system areas. Unlike other references \cite{Ohura.1990,Sauhats.2017}, we allow splitting for a broad range of generator angle differences instead of relying on a single threshold value. Finally, our splitting scheme does not depend on large amounts of training data and its decisions are fully traceable, which is favorable compared to AI-based OOS detection \cite{Senroy.2006,Paul.2020}. 

\section{Controlled Separation Scheme Overview}
\label{sec:ovrwv}
\subsection{A High-Level Description} 
\begin{figure}[t]
\centering
\includegraphics[width=0.489\textwidth]{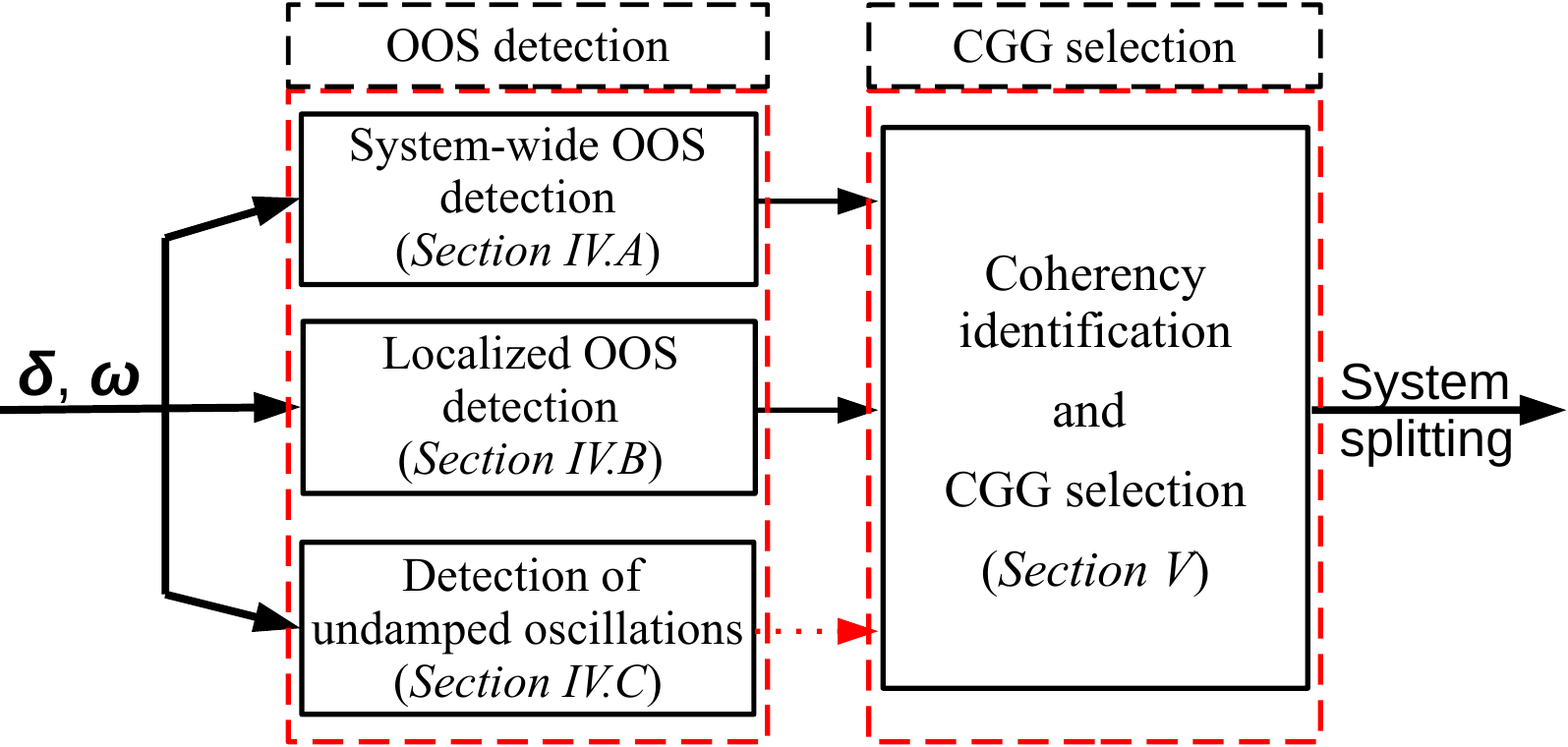}
\caption{High-level overview of the controlled splitting scheme}
\label{fig:scheme}
\end{figure}

As mentioned in the Introduction, a complete controlled system separation scheme should at least address the OOS detection and splitting boundary selection problems. In our protection scheme, the first problem is solved by proposing three OOS detection methods each of which has advantages in detecting different types of OOS conditions. These methods are marked by the red dashed line in Figure~\ref{fig:scheme}. For an $m$-generator power system, $\vect{\delta}^{m\times 1}$ is the vector of generator angles and $\vect{\omega}^{m\times 1}$ is the vector of generator frequencies; these two signal vectors are the only inputs to the three OOS detection methods. 

The second problem is tackled by defining a large number of candidate generator groupings (CGGs) that cover the most probable or critical system splitting scenarios. Once an OOS is recognized, the generator angle pattern around the moment of OOS detection is used identify the \emph{coherent machine group} that is about to lose synchronism. If the identified machine group coincides with one of the CGGs, the corresponding splitting scenario is enabled by opening the precomputed transmission lines to separate the runaway generators. The above ideas are concisely illustrated in Figure \ref{fig:scheme}.

\subsection{Selection of Input Signals}
\label{sec:inputs}
The inputs of the controlled splitting scheme, which are generically referred to as generator angles and frequencies, can have multiple realizations. For a system based on conventional synchronous generators, machines' rotor angles and speeds are preferred. They can be measured and sent as PMU analog outputs \cite{Zweigle.2013} or estimated from the standard PMU signals \cite{Ghahremani.2016}. 

In this work, we are using the angle of the voltage behind the synchronous machine's transient reactance (i.e., the angle of the machine's transient emf) as generator angle and electrical frequency at the machine terminals as generator frequency. These quantities can be straightforwardly computed from the standard PMU signals by the Ohm's law. Moreover, this choice of generator angles and frequencies allows for an easy integration of IBRs (inverter-based resources) into the proposed splitting scheme, as the recent publications on the transient stability of multi-machine multi-converter power systems suggest a unified scheme for the modeling of grid-forming IBRs and synchronous generators that is based on dynamic voltage behind transient impedance \cite{Pico.2019,Roy.2022}.

\subsection{Candidate Generator Groupings and Where to Split}
\label{sec:cgg}
Following a disturbance, generators form a number of groups that maintain approximately constant angle differences during the swings. These groups are known as \emph{coherent groups of machines}, and the loss of synchronism normally occurs along their boundaries. While there are multiple methods to retrieve generator coherency from generator angles or  frequencies \cite{naglic.2019,Sauhats.2017}, knowing the machines that go out-of-step does not automatically produce the set of transmission lines isolating those machines from the rest of the grid. In fact, computing such lines is an NP-hard constrained graph partitioning problem \cite{Tyuryukanov.2018b}. Due to this, we augment the real-time generator grouping with multiple predefined CGGs equipped with their respective splitting cutsets forming a lookup table, which could also be updated at fixed time periods based on real-time grid connectivity information provided by SCADA.

The CGGs can be defined based on offline analysis and operational experience. This is, in essence, similar to the selection of potential network splitting scenarios, an important procedure in SIPS design \cite{Sauhats.2017,Denys.1992,Ohura.1990}. A well-known offline analysis approach to the identification of generator groupings is \emph{slow coherency} \cite{chow.2013,Tyuryukanov.2020}. It utilizes the machine and transmission system data required for RMS-type simulations to return strongly coupled generator groups of various sizes. In practice, CGGs are often defined based on operational experience (e.g., in \cite{Ohura.1990}) or by observing the simulated rotor angle dynamics for multiple disturbances \cite{Denys.1992}. In some grids, the potential CGGs are obvious from the network structure \cite{Wilson.2013,Sauhats.2017}. Eventually, it is possible to use a combination of multiple methods to define the CGGs.

The use of predefined CGGs allows a seamless integration of complex optimization-based techniques for splitting boundary computation \cite{Trodden.2014}. This is because all CGGs can be optimized online in parallel for several minutes to satisfy the static power system constraints, and once an OOS occurs, the CGG that most closely matches the current unstable dynamics will be selected, thus respecting both static and dynamic system conditions. Although the coupling between the proposed OOS detection and sophisticated splitting boundary search methods was important to mention, this direction is not further pursued. Instead, the robust polynomial-time heuristic \cite{Tyuryukanov.2018b} is used in this paper to compute the splitting boundaries for each CGG. For power systems that do not support adaptive splitting boundary computations in their control centers, it is also possible to define fixed splitting boundaries for each CGG.   

\section{Rotor Angle Stability Essentials}
\label{sec:trastess}
Before considering the OOS detection problem, it is important to realize the physical foundations behind the OOS phenomena. Loss of synchronism between generators in the grid is classified as rotor angle instability (see \cite{Kundur.2004}), which can be subdivided into small-signal angle instability and transient instability. In normal operating condition, the power system operates around a \emph{stable equilibrium point} (SEP). Through a sequence of events, the equilibrium point may drift towards instability and eventually become small-signal unstable. Such situation is characterized by positive real parts of some eigenvalues of the linearized system state matrix.

\begin{figure}[t]
\centering
\includegraphics[width=0.481\textwidth]{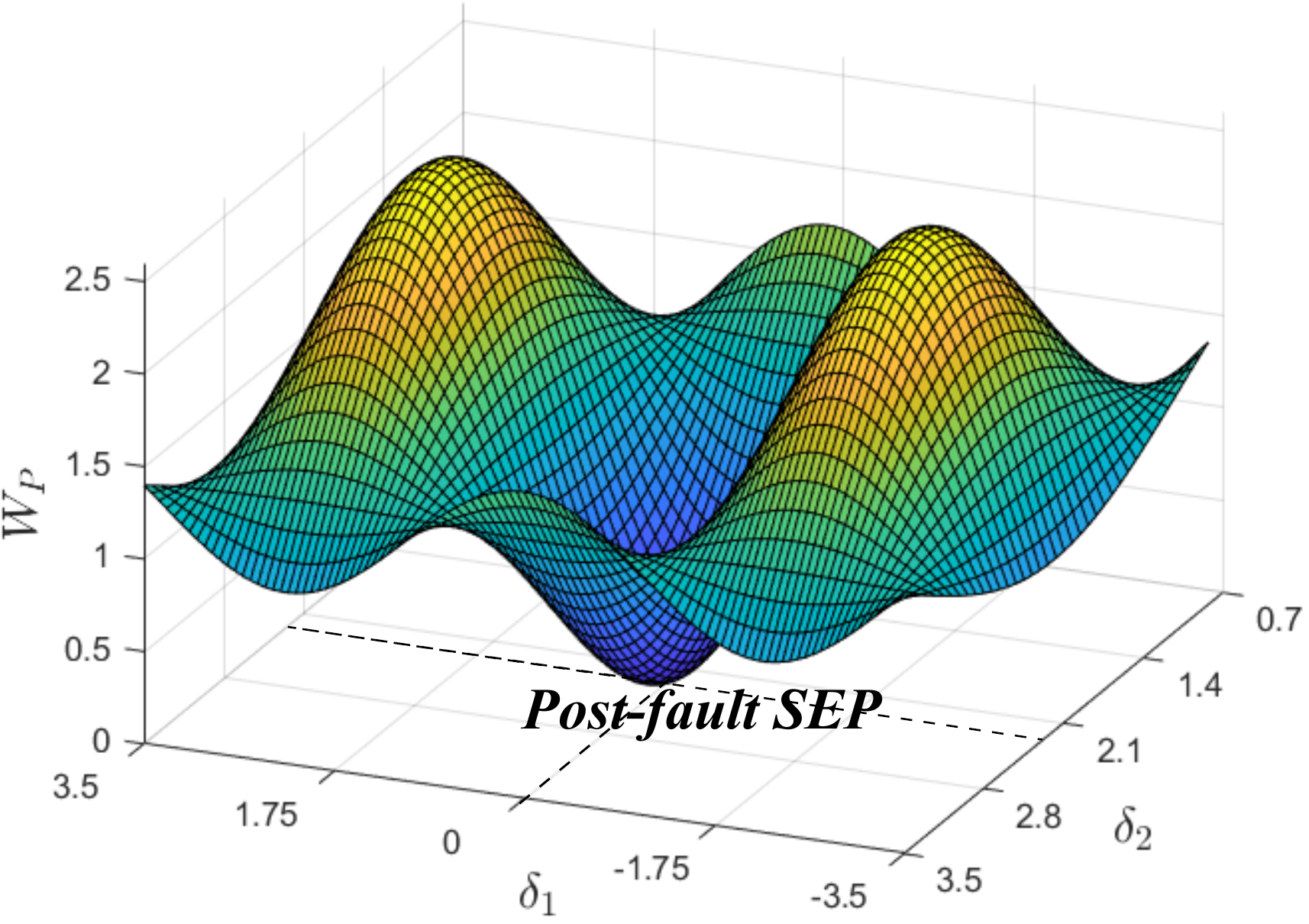}
\caption{Concept of post-fault SEP in potential energy well}
\label{fig:SEP}
\end{figure}

The transient angle instability may arise after the system is subjected to a large disturbance such as a short circuit fault. In this situation, both pre-fault and post-fault equilibria may be stable, but the system may still loose stability by escaping the \emph{region of attraction} (also known as \emph{region of stability}) of the post-fault SEP). The stable equilibria and their region of attraction are often described by the popular \emph{potential well} (or "the ball and the bowl") analogy \cite{Chiang.2011,Pai.1989}. A conceptual illustration of a post-fault SEP at the bottom of a potential energy well is shown in Figure \ref{fig:SEP}. As it can be observed from this figure, the energy needed to leave the region of attraction depends on the direction of the system trajectory. 

When a fault occurs in the system, it increases the total system energy by causing generator rotor speeds and voltage angle differences across the network to grow. The total system energy $W$ generally consists of kinetic energy $W_K$ and potential energy $W_P$:
\begin{equation}
W(\vect{\delta},~\vect{\omega},~\vect{V},~\vect{E_{fd}}) = W_K(\vect{\omega}) + W_P(\vect{\delta},~\vect{V},~\vect{E_{fd}})
\label{eqn:Wtot}
\end{equation} 

where $\vect{V}$ and $\vect{E_{fd}}$ are the vectors of bus voltages and generator field voltages. The total transient energy in \eqref{eqn:Wtot} grows during the fault and reaches a peak value at the instance of fault clearing. After the fault is removed, $W$ relative to the pre-fault or post-fault SEP is non-increasing both when the synchronism is preserved or lost \cite{Pai.1989,Chiang.2011,Padiyar.2013,Rimorov.2018}. The instant drop of generator accelerating torques after the fault clearing causes the kinetic energy to decrease and to be converted into the potential energy. As the kinetic energy decreases, the potential energy increases to confine the slow decline of the total energy $W$ caused by system damping \cite{Pai.1989,Chiang.2011,Padiyar.2013}. 


Power system can loose stability when it acquires enough total energy $W$ to leave the region of stability around the post-fault SEP. The region of stability of a SEP is delimited by the stable manifolds of the \emph{unstable equilibrium points} (UEP) that surround the SEP \cite{Chiang.2011}. Because of this, the potential energy value at the \emph{controlling unstable equilibrium point} (CUEP) $\vect{\delta^{u}}$ is essential to accurately assess the stability for a given power system initial condition and disturbance \cite{Chiang.2011}. For a SMIB system, the system trajectory is characterized by a single generator angle, which results in unique solutions for $\delta^{u}$ and the \emph{critical energy} $W_P(\delta^{u})$, both of which can be computed with the EAC. The trajectory of a multi-machine power system is characterized by $m$ rotor angles and has a large number of possible UEPs. Finding the CUEP for a given fault and its critical energy becomes a highly non-trivial task. The related complexities can be subdivided into two categories:
\begin{enumerate}
\item Computing the CUEP requires the knowledge of fault location and the real-time power system model. Additionally, even the most advanced algorithms for computing CUEP may return wrong results in certain corner cases.
\item Computing the system potential energy not only requires the availability of the real-time system model, but even the analytic expression for potential energy may not be precisely known for complex power system models \cite{Rimorov.2018}. 
\end{enumerate}

Because of the above limitations, direct stability analysis methods cannot be utilized to the full extent in a practical OOS protection scheme. Indeed, the applications of these methods are more related to contingency screening \cite{Chiang.2011,Rimorov.2018}.

\section{OOS Detection}
\label{sec:oosdetect}
\subsection{OOS Detection using COI-based Stability Indicators}
\label{sec:WkeGMM}
The proposed wide-area OOS protection scheme relies on some major attributes that are accompanying loss of synchronism to promptly determine the OOS condition in the grid. First, it is well-known \cite{Ohura.1990,Denys.1992,Sauhats.2017} that the maximal difference among generator angles should become large enough to constitute a risk to the synchronous grid operation: 
\begin{equation}
\delta^{\mathrm{max}}_{ij}[n] > \delta_{arm}
\label{eqn:minangdiff}
\end{equation}

where $\delta^{\mathrm{max}}_{ij} := \max_{\substack{1\leq i \leq m \\ 1\leq j \leq m}} |\delta_i-\delta_j|$, $n$ is the current PMU sample number, and $\delta_{arm}$ is the minimal maximal angle difference that could result in an unstable oscillation. For the simplest SMIB system, transient stability is guaranteed for $\delta\leq 90^{\circ}$; thus, $\delta_{arm}$ higher than $90^{\circ}$ is required to ensure the protection blocking for stable swings. The dynamics of multi-generator meshed power grids are more complex, and a system-specific value of $\delta_{arm}$ can be chosen instead \cite{Ohura.1990} (e.g., around $120^{\circ}$ in \cite{Ohura.1990,Sauhats.2017}). However, with our approach, $\delta_{arm}$ can be set to a minimal viable value because, unlike in \cite{Ohura.1990,Denys.1992,Sauhats.2017}, \eqref{eqn:minangdiff} is not the only transient instability criterion.

The next instability condition requires the maximal angle difference to be growing for some period of time prior to the instant of instability detection:
\begin{equation}
\delta^{\mathrm{max}}_{ij}[k]-\delta^{\mathrm{max}}_{ij}[k-1] > \varepsilon_\delta, ~ k = n-N,\ldots,n
\label{eqn:grwangdiff}
\end{equation}

where $\varepsilon_\delta$ is a small positive number (e.g., $1^{\circ}/\mathrm{s}$) and $N T_s$ is the window length for OOS detection with $T_s$ being the PMU sampling period. Clearly, the condition in \eqref{eqn:grwangdiff} is meaningful but largely superfluous if combined solely with \eqref{eqn:minangdiff}. However, it becomes more useful when combined with the subsequent conditions based on the contents of Section \ref{sec:trastess}. 

Following a disturbance, the kinetic and potential energies are oscillating in a counterphase to satisfy a certain rate of decay of the total system energy that depends on the amount of damping  present in the system \cite{Rimorov.2018}. As mentioned in Section \ref{sec:trastess}, a closed-form expression for the potential energy is hard to derive for a realistic power system. In contrary, the kinetic energy can be simply computed as \cite{Chiang.2011,Padiyar.2013}: 
\begin{subequations}
\begin{align}
&W_{K,i} = 0.5 M_i \widetilde{\omega}^2_i,~i=1,\ldots,m \label{eqn:Wki} \\
&W_K = \frac{1}{m}\sum_{i=1}^m W_{K,i} \label{eqn:WkS}
\end{align}
\label{eqn:Wk}
\end{subequations} 

where $W_{K,i}$ is the kinetic energy of generator $i$, $M_i$ is the inertia of $i$ generator, and $\widetilde{\omega}_i$ is the frequency of $i$ generator relative to the center of inertia (COI) of the AC power system containing $i$ generator. For system before splitting, the generator angles and frequencies relative to the COI are computed as follows:
\begin{equation}
\widetilde{\delta}_i = \delta_i - \frac{\sum_{i=1}^m M_i \delta_i}{\sum_{i=1}^m M_i},~\widetilde{\omega}_i = \omega_i - \frac{\sum_{i=1}^m M_i \omega_i}{\sum_{i=1}^m M_i},~\forall i
\label{eqn:coi}
\end{equation}

Furthermore, a loss of synchronism is characterized by a significant increase of the system kinetic energy, which directly follows from its definition (i.e., the presence of multiple distinct COIs in a connected AC grid). After fault clearing, the kinetic energy decreases until it reaches a local minimum, unless the fault clearing time noticeably exceeds the critical clearing time (CCT). Clearly, as loss of synchronism is characterized by the growing kinetic energy, the kinetic energy decrease can be used for protection blocking. Alternatively, we can introduce the following OOS detection condition:
\begin{subequations}
\begin{align}
&\Delta(W_K)[k] > \varepsilon_W, ~ k = n-N,\ldots,n \label{eqn:grwwke} \\
&\frac{\Delta(W_K)[k]}{\Delta(W_K)[k-1]} > \alpha_W, ~ k = n-N,\ldots,n \label{eqn:accwke}
\end{align}
\label{eqn:grwwkediff}
\end{subequations}
\indent where $\Delta(W_K)[k]:=W_K[k]-W_K[k-1]$, $\varepsilon_W$ is a small positive number and $\alpha_{W}>1$. The condition in \eqref{eqn:accwke} complements \eqref{eqn:grwwke}, as $W_K$ initially slowly grows from its local minima and this growth accelerates over time. For unstable transients, the growth acceleration of $W_K$ can be very significant. In our work, we have used $\alpha_W\in[1.05,~1.15]$, with lower values leading to higher protection sensitivity and higher values reducing the chance of false alarms. 

In some situations, using conditions \eqref{eqn:minangdiff}, \eqref{eqn:grwangdiff}, \eqref{eqn:grwwkediff} may delay the OOS detection. In particular, we have observed that for some transients $W_K(t)$ may vary very little around a local minimum (i.e., "flatten out"), thus delaying the activation of \eqref{eqn:grwwkediff}. To speed up the OOS detection for such transients, a COI-based instability indicator is introduced as follows:
\begin{subequations}
\begin{align}
&\gamma_i(t) = \widetilde{\delta}_i(t)\widetilde{\omega}_i(t),~i=1,\ldots,m \label{eqn:gammai} \\
&\gamma(t) = \frac{1}{m} \sum_{i=1}^m \gamma_i(t) \label{eqn:gammaS}
\end{align}
\label{eqn:gamma}
\end{subequations}

The meaning of \eqref{eqn:gamma} is based on the idea that loss of synchronism is accompanied by large deviations of some generator angles away from the COI. Large COI generator angles are only positively contributing to \eqref{eqn:gammaS} if the respective generator frequencies have the same sign (e.g., positive COI frequencies for large positive $\widetilde{\delta}_i$). Thus, positive values of \eqref{eqn:gammaS} indicate the power swing stages when generators tend to separate, while negative values of \eqref{eqn:gammaS} indicate the stages when generators tend to synchronize. Based on \eqref{eqn:gamma}, the following condition can be introduced to complement  \eqref{eqn:minangdiff}--\eqref{eqn:grwangdiff}:
\begin{subequations}
\begin{align}
&\gamma[k] > 0, ~ k = n-N,\ldots,n\label{eqn:posgmm} \\  
&\Delta(\gamma)[k] > \varepsilon_{\gamma}, ~ k = n-N,\ldots,n\label{eqn:grwgmm} \\
&\frac{\Delta(\gamma)[k]}{\Delta(\gamma)[k-1]} > \alpha_{\gamma}, ~ k = n-N,\ldots,n\label{eqn:accgmm}
\end{align}
\label{eqn:grwgmmdiff}
\end{subequations}

where $\Delta(\gamma)[k]:=\gamma[k]-\gamma[k-1]$, $\varepsilon_{\gamma}$ is a small positive number and $\alpha_{W}>1$. Similarly to \eqref{eqn:grwwkediff}, the tendency towards loss of synchronism is characterized by growing $\gamma$ and this growth accelerating. The used values of $\alpha_{\gamma}$ also belong to the interval $[1.05,~1.15]$. As explained above, \eqref{eqn:grwgmmdiff} was originally intended for situations with $\Delta(W_K)\approx 0$ over multiple PMU samples. However, our simulation studies have shown the effectiveness of condition \eqref{eqn:grwgmmdiff} in complementing \eqref{eqn:minangdiff}--\eqref{eqn:grwangdiff} even without considering  $W_K$ together with $\gamma$.

To further tighten our instability conditions, the notion of \emph{critical machine pair} is introduced as the tuple of machine indices corresponding to $\delta^{\mathrm{max}}_{ij}$: 
\begin{equation}
(i^{\mathrm{max}},j^{\mathrm{max}}) = \operatorname{arg~max}_{\substack{1\leq i \leq m-1 \\ i+1\leq j \leq m}} |\delta_i-\delta_j|
\label{eqn:critmac}
\end{equation}

As $\sum_{i=1}^m M_i\widetilde{\delta_i} \equiv 0$, $\widetilde{\delta}_{i^{\mathrm{max}}}$ and $\widetilde{\delta}_{j^{\mathrm{max}}}$ must have the opposite signs. That is, one of the generators in the critical machine pair must be accelerating relative to the COI, while the other one should be decelerating. To ensure this to be true at the OOS detection time, the following condition is stated: 
\begin{equation}
\widetilde{\omega}_{i^{\mathrm{max}}}[n]>\underline{\omega} \vee \widetilde{\omega}_{j^{\mathrm{max}}}[k]<-\underline{\omega}
\label{eqn:oosspd}
\end{equation}

where it is assumed that $\widetilde{\delta}_{i^{\mathrm{max}}}[k]>0$ and $\widetilde{\delta}_{j^{\mathrm{max}}}[k]<0$ and $\underline{\omega}$ represents the minimal frequency deviation to confirm generator movement away from the COI. In our case studies, we have used $\underline{\omega}$ around 0.003 p.u., possibly with higher values for lower $\delta^{\mathrm{max}}_{ij}$ and lower values for higher $\delta^{\mathrm{max}}_{ij}$. 

The conditions introduced in this section can be summarized into the following OOS detector:
\begin{equation}
\eqref{eqn:minangdiff} \wedge \eqref{eqn:grwangdiff} \wedge \eqref{eqn:oosspd} \wedge \left(  \eqref{eqn:grwwkediff} \vee \eqref{eqn:grwgmmdiff} \vee \delta^{\mathrm{max}}_{ij}[n] > \delta_{crt} \right) 
\label{eqn:oosdct1}
\end{equation}

where the condition $\delta^{\mathrm{max}}_{ij}[n] > \delta_{crt}$ enforces system splitting once the angular separation is higher than the maximum limit $\delta_{crt}$ allowed by the system operator. It is common to choose $\delta_{crt}$ around $180^{\circ}$ \cite{Ohura.1990,Sauhats.2017}, although the EAC does not directly translate to multi-machine power systems and the actual value of $\delta_{crt}$ can be system-dependent \cite{Ohura.1990,Venkatasubramanian.2016}. 

\subsection{OOS Detection based on Phase Portraits}
\label{sec:PP}
The OOS conditions in Section \ref{sec:WkeGMM} involve system-wide stability indices (e.g., $W_K$, $\gamma$, $\delta_{ij}^{\max}$) that largely overshadow the transient behavior of individual generators or generator groups. This deficiency may delay the OOS detection by several tens of milliseconds for some disturbances. To accelerate the OOS detection for such localized power swings, we introduce another OOS detection method based on \emph{phase portraits}. Its secondary objective is to back up the main OOS detection scheme in Section \ref{sec:WkeGMM}.

Phase portraits are commonly used in nonlinear system theory to portray the system trajectory in the \emph{phase space} (i.e., the space of system state variables) to study the system behavior and stability properties. In the power system literature, phase portraits in the coordinates of rotor angles and speeds are defined as a parametric equation in time $t$ both for single generators and generator groups:
\begin{subequations}
\begin{align}
\delta_{\mathcal{A}}(t) = \frac{\sum_{i\in\mathcal{A}} M_i \delta_i(t)}{\sum_{i\in\mathcal{A}} M_i} - \frac{\sum_{j\in\mathcal{G}\setminus\mathcal{A}} M_j \delta_j(t)}{\sum_{j\in\mathcal{G}\setminus\mathcal{A}} M_j} \\
\omega_{\mathcal{A}}(t) = \frac{\sum_{i\in\mathcal{A}} M_i \omega_i(t)}{\sum_{i\in\mathcal{A}} M_i} - \frac{\sum_{j\in\mathcal{G}\setminus\mathcal{A}} M_j \omega_j(t)}{\sum_{j\in\mathcal{G}\setminus\mathcal{A}} M_j} 
\end{align}
\label{eqn:phpt}
\end{subequations}

where $\mathcal{G}$ is the set of all generators and $\mathcal{A}$ is the generator group of interest. The phase portrait according to \eqref{eqn:phpt} can be seen as a reduction of a multi-machine system dynamics to the dynamics of a two-machine system in which the first machine aggregates all generators in $\mathcal{A}$ and the second machine aggregates the remaining generators \cite{Pavella.2000}.  

There is a significant body of work related to applications of phase portraits to power system transient stability analysis \cite{Pavella.2000,Salimian.2018}. Some of these works assume that the system looses stability once some phase portrait \eqref{eqn:phpt} exhibits a convex state trajectory. Our studies on the IEEE 39-bus test system \cite{Pai.1989} have shown that this condition is necessary, but not sufficient. 
\begin{figure}[t]
\centering
\includegraphics[width=0.33\textwidth]{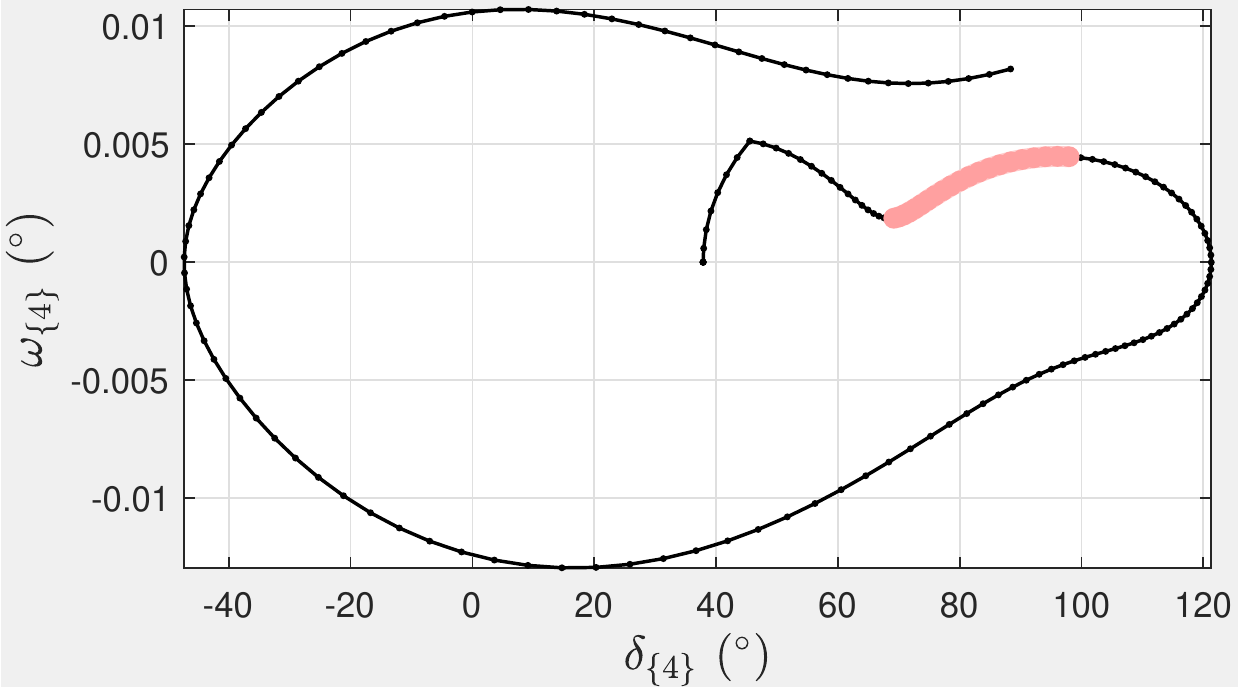}          
\caption{Phase portrait of generator 4, stable first swing}  
\label{fig:phptst}
\end{figure}
For example, Figure \ref{fig:phptst} shows how a convex trajectory (highlighted in red) may become concave again, thus resulting in a stable first swing. The shown scenario corresponds to a multiswing unstable case of an 8 cycle three-phase fault at bus 22 followed by the tripping of line 22--21 and subsequent oscillations lasting for 3.25 seconds. If the fault duration were reduced to 6 cycles, the system would remain stable, but the phase portrait of generator 4 would still be similar to Figure~\ref{fig:phptst}, although the convexity would be less pronounced. 

To improve the accuracy of OOS detection based on phase portraits, we are proposing the following set of conditions: 
\begin{subequations}
\begin{align}
& \delta_{\mathcal{A}}[n] > \delta_{arm}^{\mathrm{COI}} \label{eqn:phptdltmin}\\
& \omega_{\mathcal{A}}[k] > \omega_{\mathcal{A}}(1) \wedge \delta_{\mathcal{A}}[k] > \delta_{\mathcal{A}}(1),~k = n-N,\ldots,n \label{eqn:phptomgmin} \\
& \Delta(\delta_{\mathcal{A}})[k] > \varepsilon_{\delta}^{\mathrm{pp}},~k = n-N,\ldots,n \label{eqn:phptdltacc} \\
& \Delta(\omega_{\mathcal{A}})[k] > \varepsilon_{\omega}^{\mathrm{pp}},~k = n-N,\ldots,n  \label{eqn:phptomgacc} \\
& \frac{\Delta(\omega_{\mathcal{A}})[k]}{\Delta(\delta_{\mathcal{A}})[k]} > \varepsilon_{\frac{d\omega}{d\delta}}^{\mathrm{pp}},~k = n-N,\ldots,n 
\end{align}
\label{eqn:phptCOND}
\end{subequations}

where $\Delta(\delta_{\mathcal{A}})[k] = \delta_{\mathcal{A}}[k]-\delta_{\mathcal{A}}[k-1]$, $\Delta(\omega_{\mathcal{A}})[k] = \omega_{\mathcal{A}}[k]-\omega_{\mathcal{A}}[k-1]$, $\varepsilon_{\delta}^{\mathrm{pp}}$ and $\varepsilon_{\omega}^{\mathrm{pp}}$ are small positive numbers, and $\varepsilon_{\frac{d\omega}{d\delta}}^{\mathrm{pp}}>0$ defines the minimal slope of a diverging phase portrait \eqref{eqn:phpt}. 

In \eqref{eqn:phptCOND}, conditions \eqref{eqn:phptdltmin}--\eqref{eqn:phptomgacc} are inspired by the interpretation of \eqref{eqn:phpt} as a reduction of a multi-machine system to a SMIB system \cite{Pavella.2000}. When a SMIB system passes its critical angle $\delta^u$, its rotor angle and speed start to accelerate, which is reflected by \eqref{eqn:phptdltacc}--\eqref{eqn:phptomgacc}. Additionally, an unstable trajectory is effectively constrained by \eqref{eqn:phptdltmin}--\eqref{eqn:phptomgmin} to belong to the first quadrant of the phase plane that is centered at the pre-fault SEP. The minimal critical angle $\delta_{arm}^{COI}$ in \eqref{eqn:phptdltmin} should be smaller than $\delta_{arm}$ in \eqref{eqn:minangdiff} at around $100^{\circ}$ as $\delta_{\mathcal{A}}$ represents an angle difference between COIs of two generator groups rather than a pair of individual machines. 

The global system stability index $W_K$ must be added to \eqref{eqn:phptCOND} to validate the unstable behavior of an individual phase portrait. By considering this, the following conservative instability indicator can be proposed:
\begin{equation}
\eqref{eqn:minangdiff} \wedge \eqref{eqn:grwangdiff} \wedge \eqref{eqn:phptCOND} \wedge \left(\eqref{eqn:grwwke} \vee \delta_{\mathcal{A}}[n]\geq 180^{\circ} \right)
\label{eqn:oosdct2}
\end{equation}

where the condition $\delta_{\mathcal{A}}[n]\geq 180^{\circ}$ is based on the previously mentioned analogy of  \eqref{eqn:phpt} and SMIB. 
 
Clearly, if \eqref{eqn:oosdct2} holds for any generator, the whole system can be considered as unstable, and the actual splitting grouping can be decided by using the methodology of Section \ref{sec:sss}.

\subsection{Detection of Undamped Power Swings}
\label{sec:Wke}
Undamped electromechanical oscillations constitute a separate type of unstable power swings. Transient instabilities arising after clearing of severe short circuit faults typically result in one or several periods of growing oscillations (i.e., single-swing or multi-swing instabilities \cite{Pavella.2000}) that culminate in an unbounded angle separation between some parts of the system. In contrast, undamped oscillations caused by a change in system operating condition or topology may include many cycles that grow over several tens of seconds (e.g., the 1996 western USA/Canada blackout \cite{Kundu.2016}). Such slowly evolving instabilities have received a special treatment in the literature with several detection methods proposed so far (e.g., \cite{Sun.2011,Salimian.2018}). 

The presence of a negatively damped electromechanical oscillation mode must be characterized by growing angles and frequencies of some generators. Consequently, the kinetic energy \eqref{eqn:Wki} of certain generators should exhibit oscillations of growing amplitude. If the number of growing peaks of $W_{K,i}(t)$ surpasses the threshold $\hat{\kappa}$ for some generator $i$, an undamped oscillation can be registered. The value of $\hat{\kappa}$ can be tailored to the typical dynamics of a particular power system. However, it should be set to higher than 4 to separate the detection of multiswing transient instability and undamped oscillations. Unlike the system kinetic energy \eqref{eqn:WkS}, the individual generator kinetic energies are guaranteed to reach the absolute minimum of zero within each oscillation cycle. This feature makes the tracking of local maxima of $W_{K,i}$ more simple and reliable. 
\begin{figure}[t]
\centering
\includegraphics[width=0.48\textwidth]{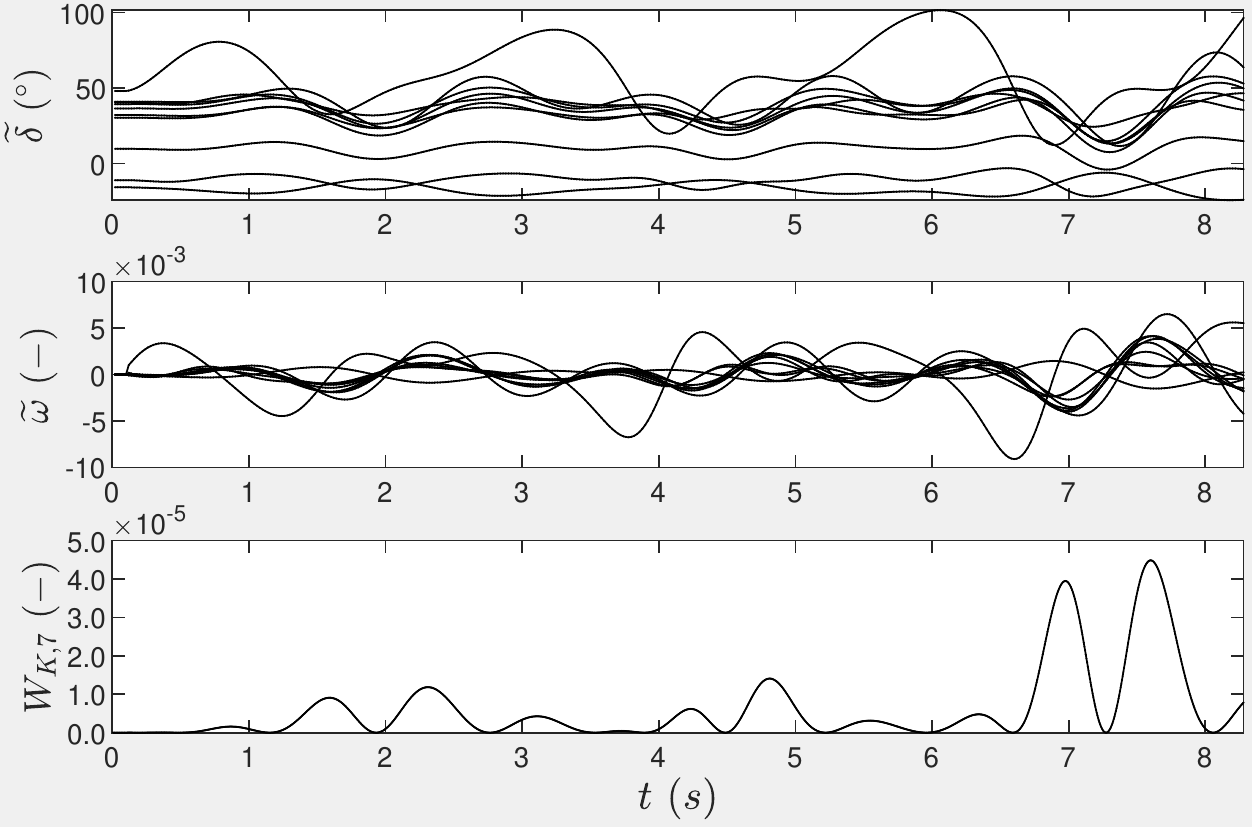}          
\caption{Undamped oscillation after disconnecting of line 28--29}
\label{fig:undpd}
\end{figure}

The above concept can be best illustrated by a case study. In the IEEE 39-bus test system, consider line 28--29 to be switched off at $t=0.1~s$ without a prior fault. The resulting transient can be seen in Figure \ref{fig:undpd}. The bottom plot in Figure \ref{fig:undpd} illustrates the five growing peaks of the kinetic energy of generator 7 at bus 36. The first peak is counted at $t\approx 1.6$~s as the local maxima of $W_{K,i}$ corresponding to very low values of $\widetilde{\omega}_i$ can be neglected. The counter of $W_{K,i}$ maxima $\kappa_i$ is increased each time the current local maximum of $W_{K,i}$ exceeds the value of the previous absolute maximum of $W_{K,i}$ by 5~\%. Once $\kappa_i$ reaches $\hat{\kappa}\geq 5$, undamped oscillations can be confirmed. However, undamped oscillations do not always lead to a loss of synchronism, as they can also be timely mitigated by specialized damping control schemes. Thus, OOS is still detected by \eqref{eqn:oosdct1} and \eqref{eqn:oosdct2}, while the detected presence of undamped oscillations prior to the OOS detection could be used to customize the splitting decisions (see Section \ref{sec:conclus}).

\section{Generator Coherency and System Splitting}
\label{sec:sss}
Upon OOS detection, groups of generators moving together away from the COI should be identified. Next, the identified runaway generators are matched with the predetermined CGGs to obtain the branches to be opened to separate these generators from the rest of the grid. These two steps constitute our answer to the "where" question of controlled splitting, which has the two reasons. First, the NP-hard splitting boundary search cannot be reliably solved within a few milliseconds after OOS detection to conform to the ongoing OOS transient. Second, known online generator coherency estimation methods may be prone to errors both due to algorithmic  imperfections and ambiguities in the input data (e.g., two distant generators might have similar swings over a time window). 

A common approach for coherency estimation relies on correlation coefficients \cite{Sauhats.2017,Salimian.2018}, but it neglects the magnitude information that could be highly valuable when generator angles diverge during OOS. Large COI generator angle magnitudes indicate a higher probability of the corresponding generators to loose synchronism, which also explains the lower importance of generator frequencies for generator grouping during OOS. To capture both the angle movement pattern and magnitude, we use COI generator angles relative to their prefault (quasi-steady state) values $\widetilde{\delta}_{0,i}$ as signals for coherency estimation: $\Delta\widetilde{\delta}_i =\widetilde{\delta}_i - \widetilde{\delta}_{0,i}$. 

\begin{algorithm}[t]
\begin{algorithmic}[1]
\renewcommand{\algorithmiccomment}[1]{\small//~#1\normalsize} 
\Statex \textbf{Input:} $\Delta\widehat{\delta}$, set of CGGs
\Statex \textbf{Output:} Critical generator bipartitioning $\Pi$
\State $\mathcal{C}\gets \emptyset$;~$\Delta\widehat{\delta}^{\mathrm{max}}_{ij}\gets \matr{0}_{1\times k}$; \label{alg:cndgrp0}
\For{$k=1,\ldots,h$}
\State $\Delta\widehat{\delta}^{\mathrm{max}}_{ij}[k]\gets \max_{1 \leq g \leq m}(\Delta\widehat{\delta}_g[k])$
\State $\vect{ord}\gets \mathrm{Descending~order~of~entries~in~\Delta\widehat{\delta}_g[k]}$
\For{$g=1,\ldots,m$}
\State $\mathcal{C}\gets \mathcal{C}\cup \vect{ord}[1,\ldots,g]$ \label{alg:cndgrp}
\EndFor
\EndFor   \label{alg:cndgrp1}
\If{$\exists k=1,\ldots,h-1: \Delta\widehat{\delta}^{\mathrm{max}}_{ij}[k+1]-\Delta\widehat{\delta}^{\mathrm{max}}_{ij}[k]<0$}
\State \Return $\emptyset$ \label{alg1:chckdvrg}
\EndIf
\State $\Pi\gets\emptyset$ 
\For{$\mathcal{A}\in\mathcal{C}$}
\If{~\parbox[t]{\dimexpr\linewidth-\algorithmicindent}{$\phi_{\mathcal{A}}[k]<\phi_{\mathcal{B}}[k], \forall k=1,\ldots,h, \forall \mathcal{B}\in\mathcal{G}\setminus\mathcal{A} \wedge\\ d_{\mathcal{A}}[k+1]-d_{\mathcal{A}}[k]>0, \forall k=1,\ldots,h-1$\strut}} \label{alg:condbprt}
\State $\Pi\gets\mathcal{A}$; \textbf{break}; \label{alg:crtbprt}
\EndIf
\EndFor
\If{$\Pi=\emptyset \vee \Pi\not\in\mathrm{CGG}$}
\State \Return $\emptyset$ \label{alg1:chckpttrn}
\EndIf\\
\Return $\Pi$
\end{algorithmic}
\caption{Generator coherency following OOS detection}
\label{alg:gencoh}
\end{algorithm}

As the proposed controlled separation approach should protect against all types of OOS conditions, long observation windows for generator coherency estimation are avoided. Instead, we group the generators based on their predicted relative angles $\Delta\widehat{\delta}$, which are obtained by Taylor expansion of the $\Delta\widetilde{\delta}$ time series. We are using the Taylor prediction method proposed in \cite{Echeverria.2012} with the commonly used time horizon of 0.1 s. This corresponds to the prediction window length of $h=6$ samples for $T_s=1/60$ s. We choose to always identify two groups of generators: \emph{critical machines} CM and \emph{noncritical machines} NM \cite{Pavella.2000}. By doing so, we avoid the insecurities associated with the real-time selection of the number of clusters. After all, if an OOS condition has been detected, at least one generator group must be separated. In most of cases power systems split into two groups upon the OOS emergence \cite{Pavella.2000}. In case of a \emph{multimode} OOS, splitting can be performed sequentially if the separation of the initial CM does not stop the OOS. The detection of CM is based on the metric in \eqref{eqn:phidlt}, which is evaluated for all \emph{candidate generator bipartitionings} (CBs) $(\mathcal{A},\mathcal{G}\setminus\mathcal{A})$ for $k=1,\ldots,h$: 
\begin{subequations}
\begin{align}
d_{\mathcal{A}}[k] &= \left| \frac{\sum_{i\in\mathcal{A}}\Delta\widehat{\delta}_i[k]}{\mathrm{card(\mathcal{A})}} - \frac{\sum_{i\in\mathcal{G}\setminus\mathcal{A}}\Delta\widehat{\delta}_i[k]}{\mathrm{card}(\mathcal{G}\setminus\mathcal{A})} \right|\label{eqn:dstdlt}\\
\phi_{\mathcal{A}}[k] &=\frac{ \underset{\substack{i\in\mathcal{A} \\ j\in\mathcal{A}}}{\max} \left|\Delta\widehat{\delta}_i[k]-\Delta\widehat{\delta}_j[k]\right|+\underset{\substack{i\in\mathcal{G}\setminus\mathcal{A} \\ j\in\mathcal{G}\setminus\mathcal{A}}}{\max}\left|\Delta\widehat{\delta}_i[k]-\Delta\widehat{\delta}_j[k]\right| }{d_{\mathcal{A}}[k]} \label{eqn:phidlt}
\end{align}
\end{subequations}

The metric in \eqref{eqn:phidlt} favors machine bipartitionings that show small maximal distance within groups $\mathcal{A}$ and $\mathcal{G}\setminus\mathcal{A}$ and large distance between the groups' centroids \eqref{eqn:dstdlt}. If group $\mathcal{A}$ has the smallest value of \eqref{eqn:phidlt} over the whole prediction time window, then it could represent the CM. The complete process  for identifying CM is summarized in Algorithm \ref{alg:gencoh}. In it, lines \ref{alg:cndgrp0}--\ref{alg:cndgrp1} illustrate how the CBs $(\mathcal{A},\mathcal{G}\setminus\mathcal{A})$ are collected into set $\mathcal{C}$ via exhaustive enumeration, and the maximal angle difference $\Delta\widehat{\delta}^{\mathrm{max}}_{ij}$ is computed similarly to $\delta_{ij}^{\mathrm{max}}$ in \eqref{eqn:minangdiff} for each $k$. Next, if $\Delta\widehat{\delta}^{\mathrm{max}}_{ij}$ does not monotonically increase (a sanity check), the OOS condition is not certain, and Algorithm~\ref{alg:gencoh} terminates on line \ref{alg1:chckdvrg} for the current PMU time stamp. Otherwise, for all CBs in $\mathcal{C}$ metrics \eqref{eqn:dstdlt} and \eqref{eqn:phidlt} are evaluated to select the critical bipartitioning $\Pi=(\mathrm{CM},\mathrm{NM})$ on line \ref{alg:crtbprt}. Finally, if no CB satisfies the conditions on line \ref{alg:condbprt} or $\Pi$ does not belong to the set of protected scenarios in CGG, Algorithm~\ref{alg:gencoh} terminates on line \ref{alg1:chckpttrn} for the current PMU time stamp. Otherwise, the precomputed (or predefined) set of lines associated with $\Pi$ is opened to implement the system splitting.

\begin{figure}[t]
\centering
\includegraphics[width=0.489\textwidth]{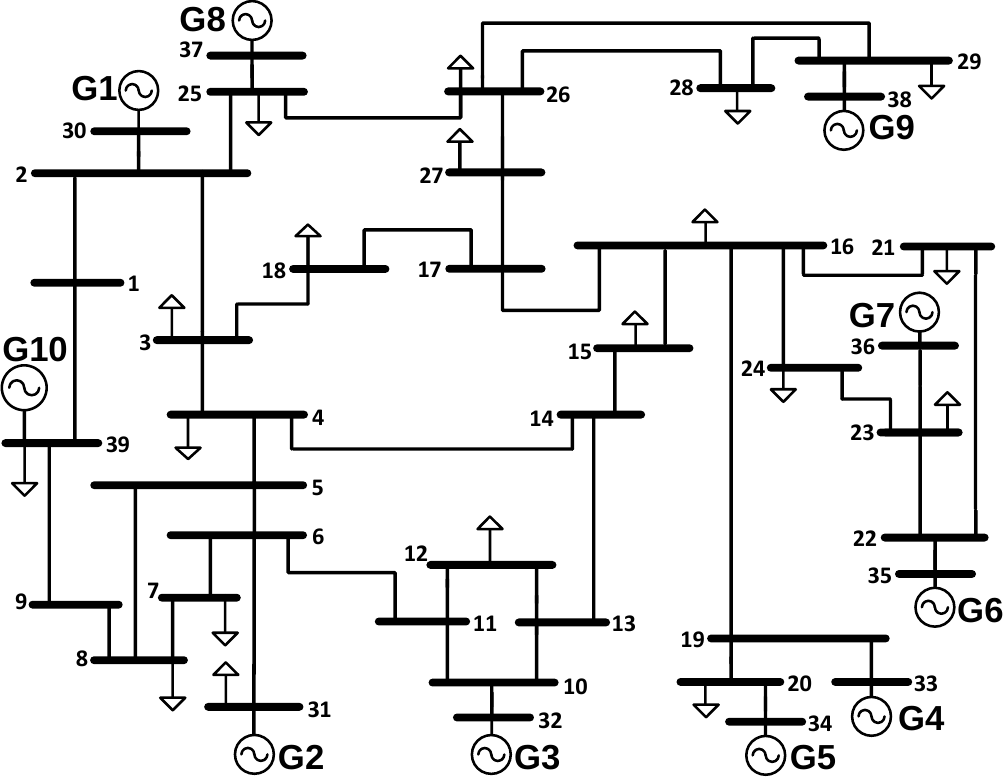}
\caption{The IEEE 39-bus test system \cite{Pai.1989}}
\label{fig:ieee39}
\end{figure}

\section{Case Studies}
\label{sec:casestud}
\subsection{Study Setup}
\label{sec:casesetup}
This section aims to illustrate the operation of the proposed controlled splitting scheme on the example of two large disturbances occurring in the IEEE 39-bus test power system. The simulations are performed with the help of the third version of the MATLAB Power System Toolbox \cite{Chow.1992,Rogers.2000}, which contains a  file with the dynamic data of the IEEE 39-bus test system as it is given in \cite{Pai.1989}. The corresponding one line diagram is shown in Figure \ref{fig:ieee39}. For the studied benchmark system, the following CGGs have been defined:
\begin{align}
&\{1,8\},~\{2,3\},~\{4,5\},~\{6,7\},~\{1,8,9\},~\{4,5,6,7\},~\{8,9\},\nonumber \\
&\{1,8,9,10\},~\{2,3,10\},~\{4,5,6,7,9\},~\{1,8,10\}
\label{eqn:cgg39}
\end{align}

In \eqref{eqn:cgg39}, each CGG is represented by the smaller of its two machine groups. The position of a CGG in \eqref{eqn:cgg39} implies its scenario number. The CGGs in \eqref{eqn:cgg39} have been found by applying the slow coherency algorithm in \cite{Tyuryukanov.2020} and the experience with OOS simulations in the test system to obtain a comprehensive set of splitting scenarios. The lines to be opened for system splitting are selected by the algorithm in \cite{Tyuryukanov.2018b} with the objective of minimal power flow disruption \cite{Kundu.2016}. 

The OOS detection performance is assessed by the maximal voltage angle difference across a network branch at  the OOS detection time $\theta^{\mathrm{max}}_{i-j}$, whereby $i-j$ is the corresponding network branch.  This is related to the common method of registering an OOS condition once the angles across some cutset in the network approach $180^{\circ}$ or other large values \cite{Padiyar.2006}. Thus, if $\theta^{\mathrm{max}}_{i-j}$ is small, the OOS detection can be considered as fast relative to the conventional methods. Additionally, we recall the popular swing centre voltage (SCV) technique \cite{PSRC.2005} that assumes OOS condition as  separation of two equivalent generators, which manifests itself by a nearly zero voltage at some point in the grid (the swing centre) when the separation becomes inevitable (i.e., the angular separation between the equivalent generators reaches \ang{180}). To quantify this OOS assessment approach, we provide the plots of bus voltage magnitudes over time. For an early OOS detection, the lowest bus voltage (the presumed SCV) should be significantly above zero at the moment of OOS detection.

The key scheme parameters have been set as follows: $T_s=1/60$ s, $N T_s = 9$ s, $\delta_{arm}=120^{\circ}$, $\delta_{arm}^{\mathrm{COI}}=100^{\circ}$, $\delta_{crt}=220^{\circ}$. 

In figures showing post-splitting transients, generator angles and frequencies are plotted relative to the COI of their respective island.

\subsection{Illustrative Case Studies on the IEEE 39-Bus Test System }
\label{sec:exmpl} 
\begin{figure}[t]
\centering
\includegraphics[width=0.48\textwidth]{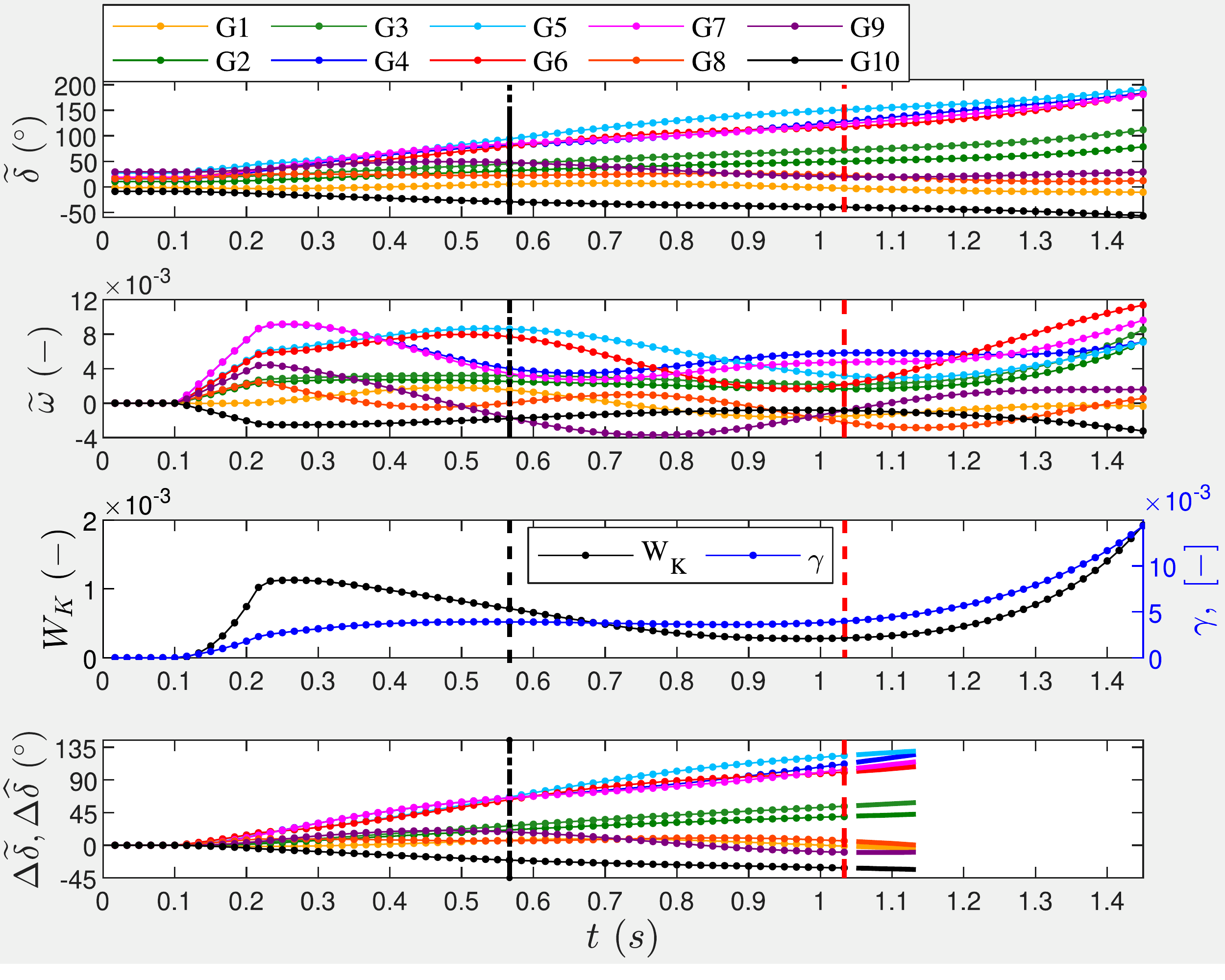}
\caption{Selected OOS detection indices for fault on line 16--17} 
\label{fig:flt1617}
\end{figure}

\begin{figure}[t]
\includegraphics[width=0.461\textwidth,left]{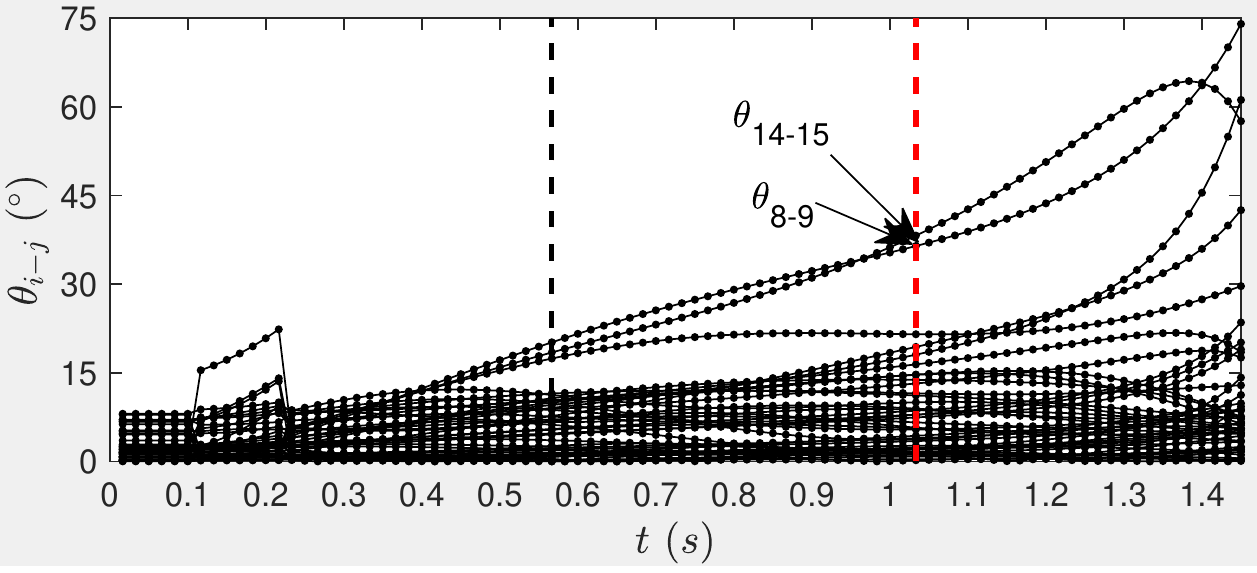}
\caption{Branch angle differences for fault on line 16--17} 
\label{fig:theta1617}
\end{figure}

\begin{figure}[t]
\includegraphics[width=0.461\textwidth,left]{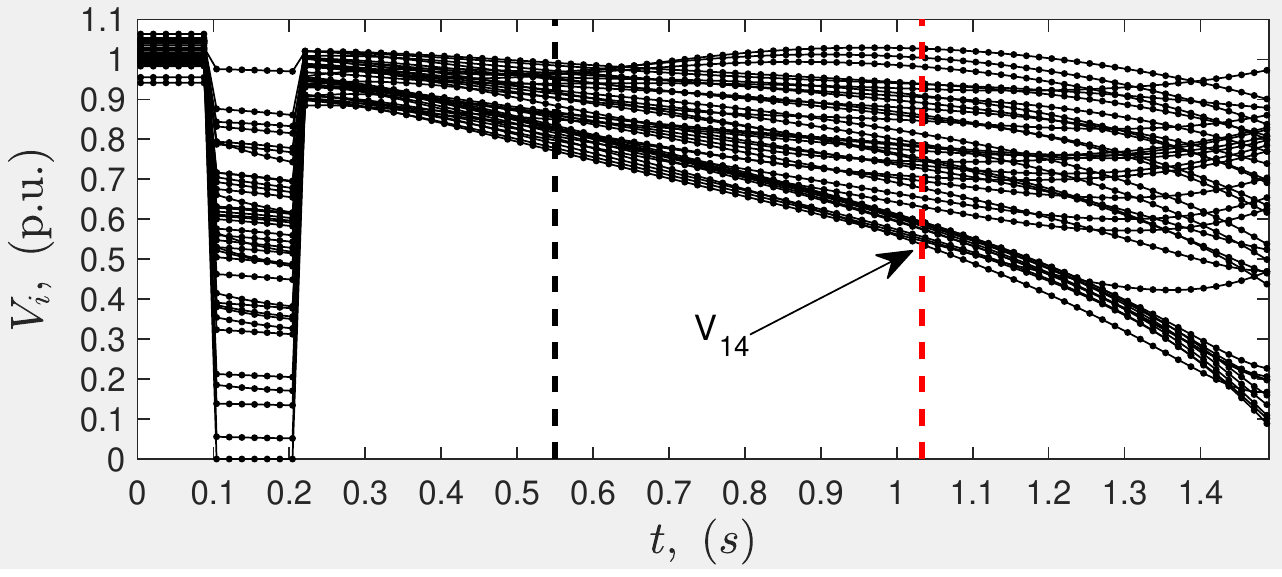}
\caption{Bus voltage magnitudes for fault on line 16--17} 
\label{fig:volt1617}
\end{figure}

As the first scenario, consider a 7-cycle (60~Hz) three phase fault occurring at $t=0.1~s$ on line \mbox{16--17} close to bus 16, which is subsequently cleared through the line tripping. The key protection signals  can be seen in Figure \ref{fig:flt1617}, in which the OOS detection time is marked by the red dashed line, while the time instant of \eqref{eqn:minangdiff} (i.e., of protection arming) is marked by the black dashed line. In the three upper plots, the shown evolution of the transient beyond the red line  is meant to confirm the instability. The lowest plot shows $\Delta\widetilde{\delta}$ and $\Delta\widehat{\delta}$ signals introduced in Section \ref{sec:sss}. In addition, the branch angle differences $\theta_{ij}$ and bus voltage magnitudes are shown in Figures \ref{fig:theta1617}--\ref{fig:volt1617} to demonstrate the OOS detection performance. 


As shown in Figure \ref{fig:flt1617}, the total kinetic energy drops after the fault clearing, but this decrease stagnates as the kinetic energy stays nearly the same from $t\approx0.9$ s, thus disabling \eqref{eqn:grwwkediff}. However, $\gamma$ starts to increase from $t\approx0.9$ s and enables \eqref{eqn:grwgmmdiff}. The final decision is made by \eqref{eqn:oosdct1} enabled by \eqref{eqn:grwgmmdiff} at $t=1.033$ s.  At that moment, $\theta^{\mathrm{max}}_{14-15}$ only equals to $38^{\circ}$, $\theta_{8-9}=36.5^{\circ}$, $\theta_{1-2}=21.5^{\circ}$ (cf. Figure \ref{fig:theta1617}), while the lowest bus voltage is 0.54 p.u. at bus 14. By applying Algorithm \ref{alg:gencoh} to the predicted values $\Delta\widehat{\delta}$  (shown on the lowest plot of Figure \ref{fig:flt1617} in the thick lines after the OOS detection), CGG~6 (i.e., $\{4,5,6,7\}$) is identified as the most suitable splitting scenario, which is consistent with the fault location. Then the system splitting is realized by disconnecting line 14--15. The comparison of the predicted generator angles with the actual ones (seen on the topmost plot of Figure \ref{fig:flt1617}) shows the maximal absolute error of $0.7^{\circ}$, maximal relative error of 1~\%, and mean relative error of 0.45~\% for the prediction horizon of 0.1~s, which is a demonstrably good accuracy. Noteworthy, generator frequencies in Figure~\ref{fig:flt1617} do not show a clear pattern that could be used to identify the splitting configuration. 
\begin{figure}[t]
\centering
\includegraphics[width=0.48\textwidth]{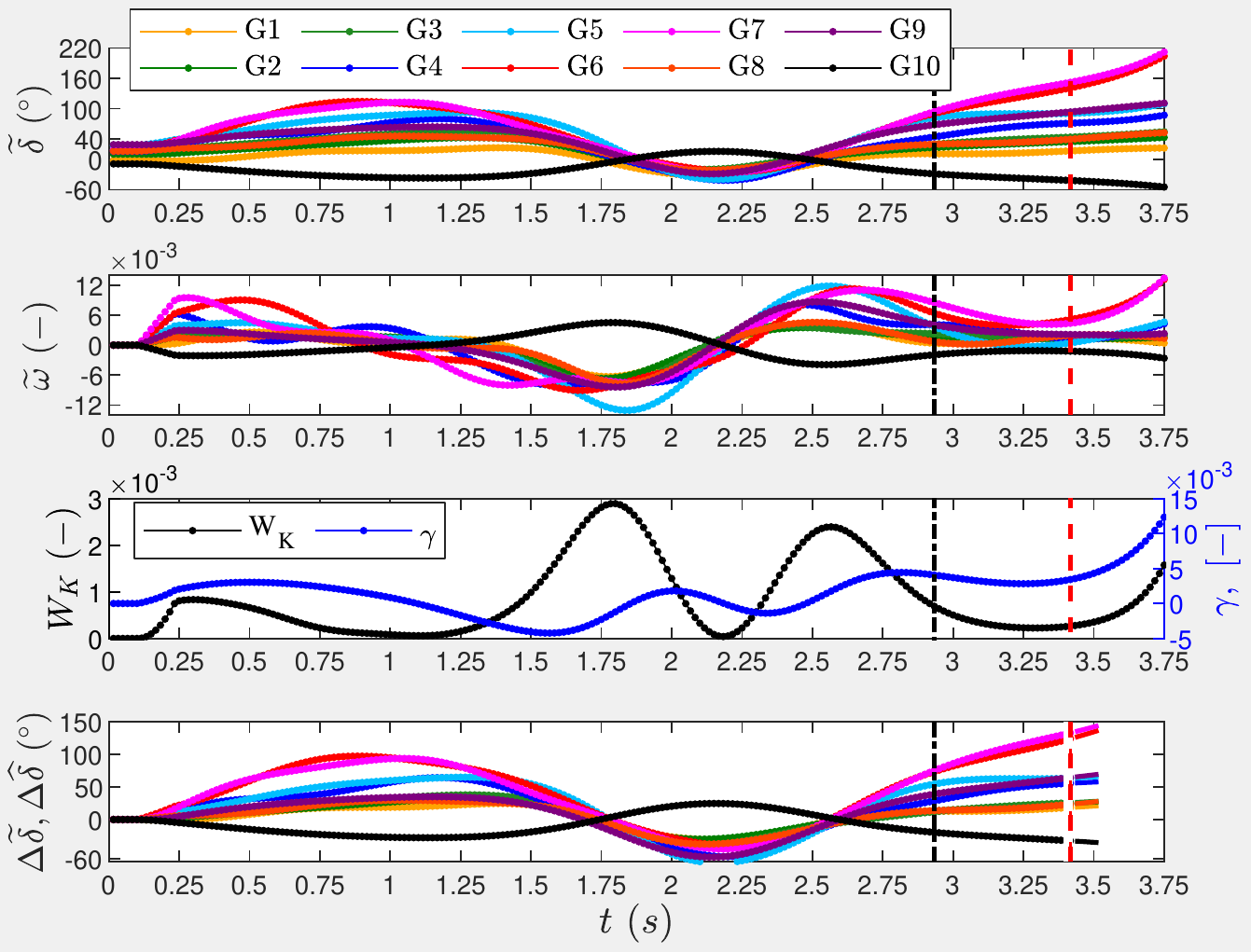}
\caption{Selected OOS detection indices for fault on line 21--22} 
\label{fig:flt2122}
\end{figure}
\begin{figure}[t]
\includegraphics[width=0.465\textwidth,left]{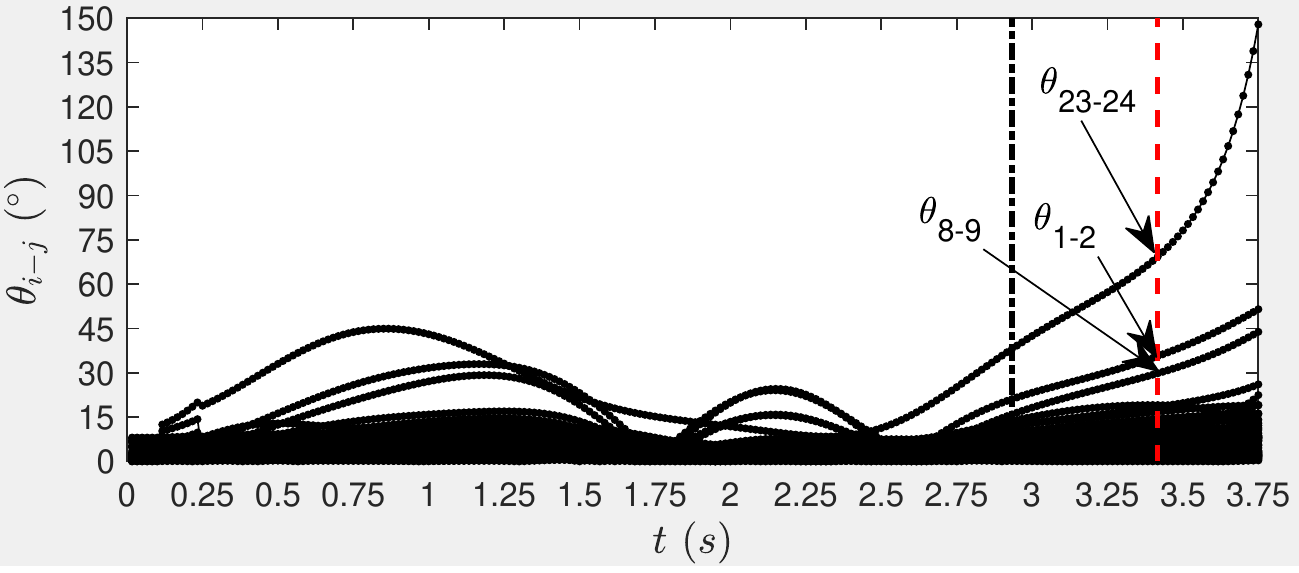}
\caption{Branch angle differences for fault on line 21--22} 
\label{fig:theta2122}
\end{figure}
\begin{figure}[t]
\includegraphics[width=0.465\textwidth,left]{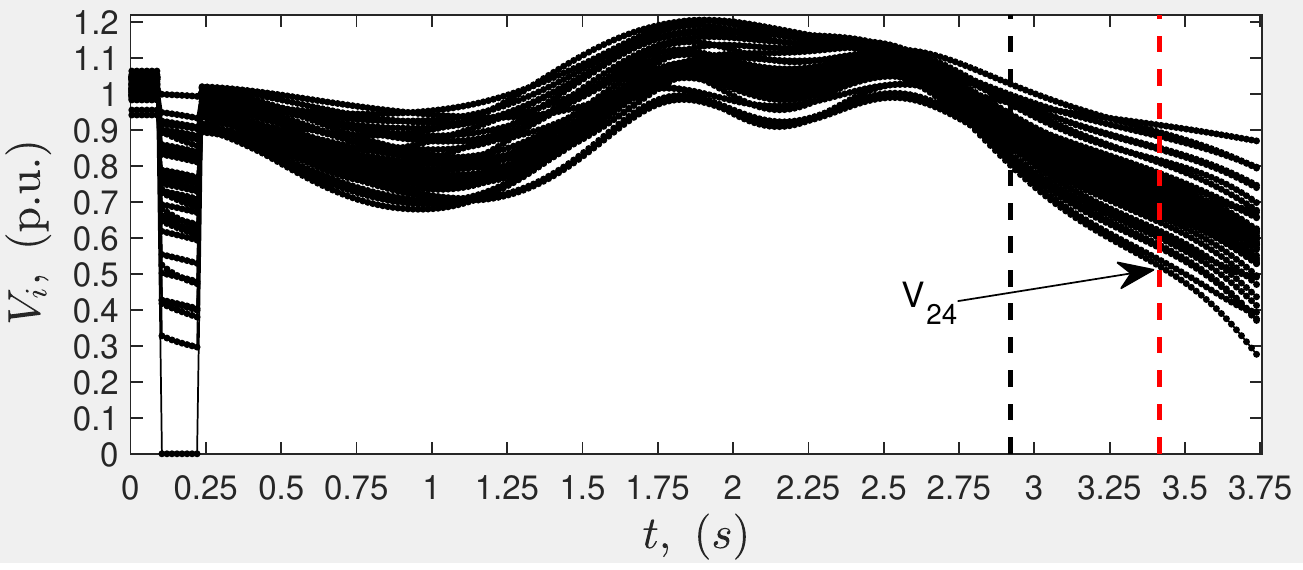}
\caption{Bus voltage magnitudes for fault on line 21--22} 
\label{fig:volt2122}
\end{figure}

\begin{figure}[t]
\centering
\includegraphics[width=0.495\textwidth]{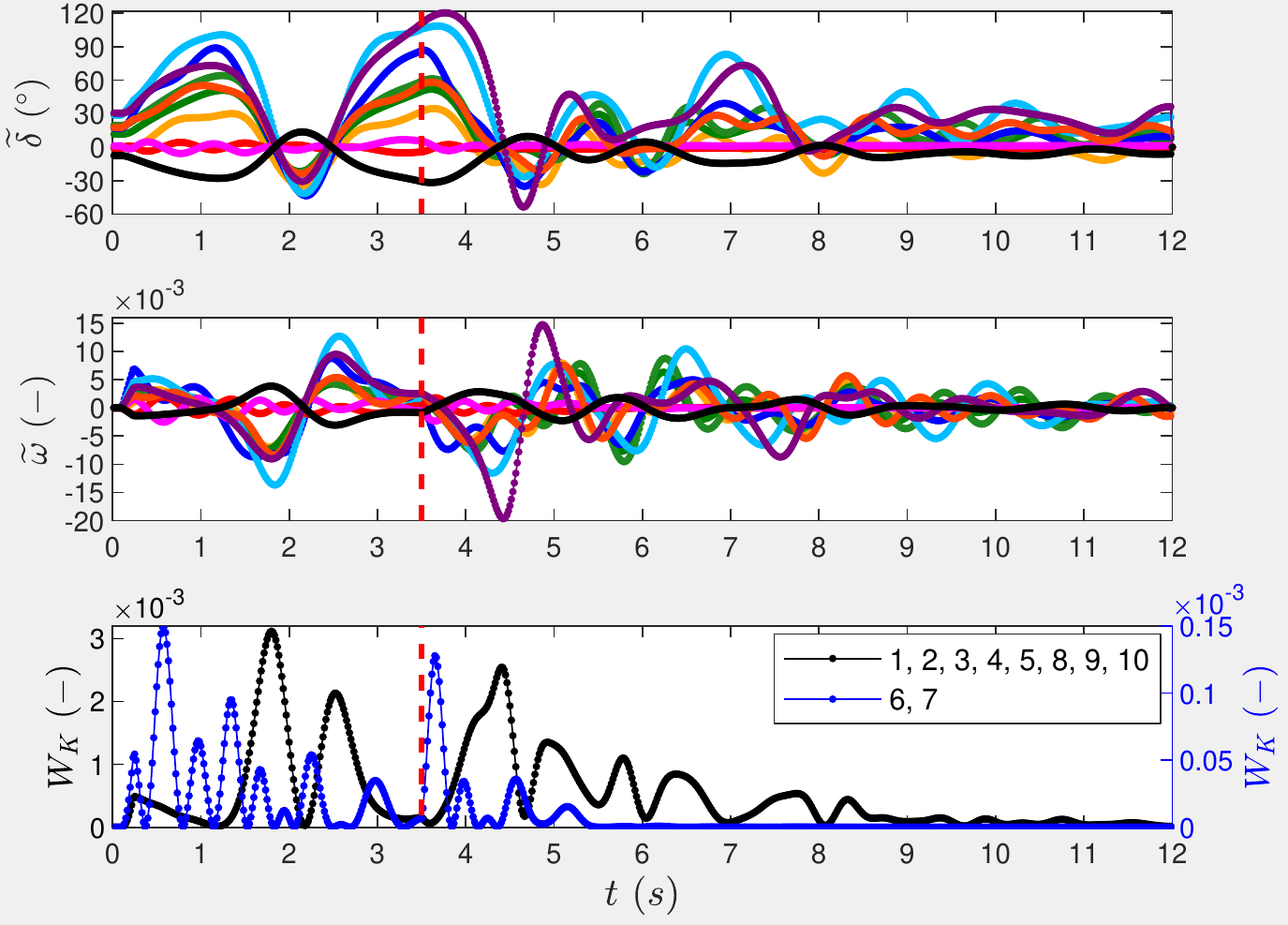}
\caption{Splitting transients after fault on line 21--22} 
\label{fig:flt2122PF}
\end{figure}
\begin{figure}[t]
\includegraphics[width=0.457\textwidth,left]{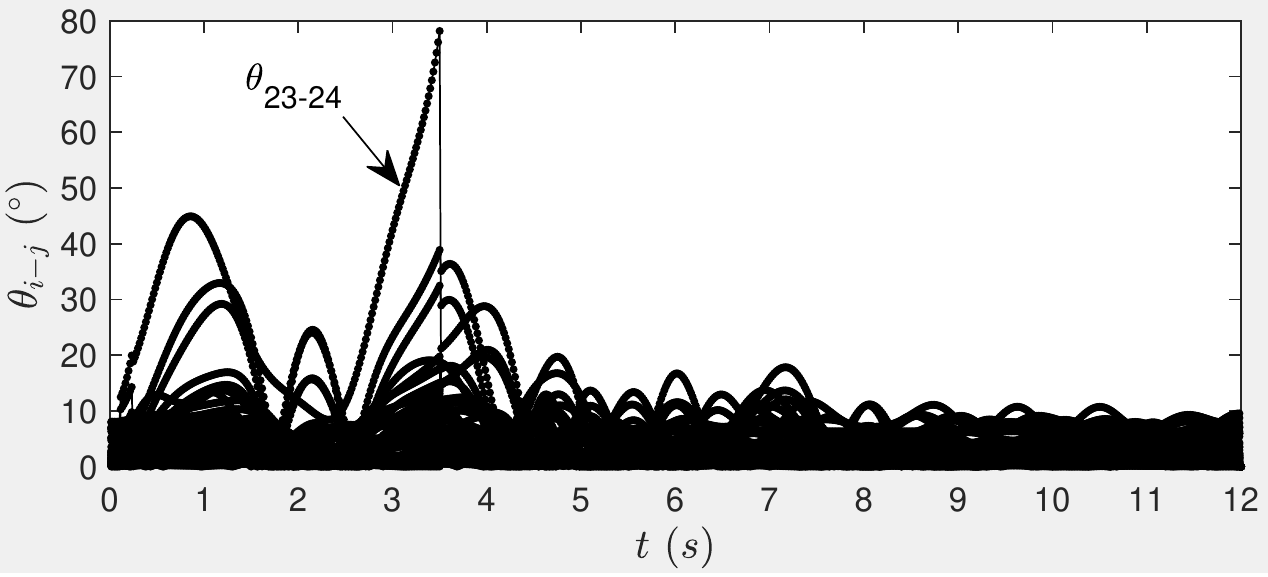}
\caption{Branch angle differences for splitting transients after fault on line 21--22}
\label{fig:theta2122PF}
\end{figure}

For an example of a multiswing instability, consider a three phase fault at $t=0.1$ s on line 21--22 close to bus 21. After an 8-cycle fault-on period, the fault is cleared by opening the line. From Figure~\ref{fig:flt2122} it can be seen that the first power swing is stable and the OOS occurs at the second swing. Similarly to the previous case, the enabling condition is \eqref{eqn:oosdct1} through \eqref{eqn:grwgmmdiff}, which is triggered by the positive and growing $\gamma$. At the moment of OOS detection, $\theta^{\mathrm{max}}_{23-24}=69^{\circ}$, $\theta_{1-2}=35^{\circ}$, $\theta_{8-9}=30^{\circ}$ and the lowest bus voltage is 0.52 p.u. at bus 24, as it can be seen in Figures \ref{fig:theta2122}--\ref{fig:volt2122}. The predicted generator angles have maximal absolute error of $0.6^{\circ}$, maximal relative error of 0.85~\%, and mean relative error of 0.35~\% for the same horizon of 0.1~s. The selected splitting scenario is CGG~4 (i.e., $\{6,7\}$), which is implemented by opening line 16--24 85 ms after the OOS detection (i.e., at $t=3.508~s$). Noteworthy, line 16--24 has been opened to limit the power imbalance in the formed islands, although its angle difference is relatively low. A local OOS protection (e.g., one based on the swing center voltage) would trip line 23--24, as it is evident from Figure \ref{fig:theta2122}. After splitting, the formed islands remain in synchronism,  which can be judged by the kinetic energies of generators in each island converging to zero (see Figures~\ref{fig:flt2122PF}--\ref{fig:theta2122PF} for the $W_K$ and $\theta_{ij}$ perspectives respectively).
\begin{figure}[t]
\centering
\includegraphics[width=0.495\textwidth]{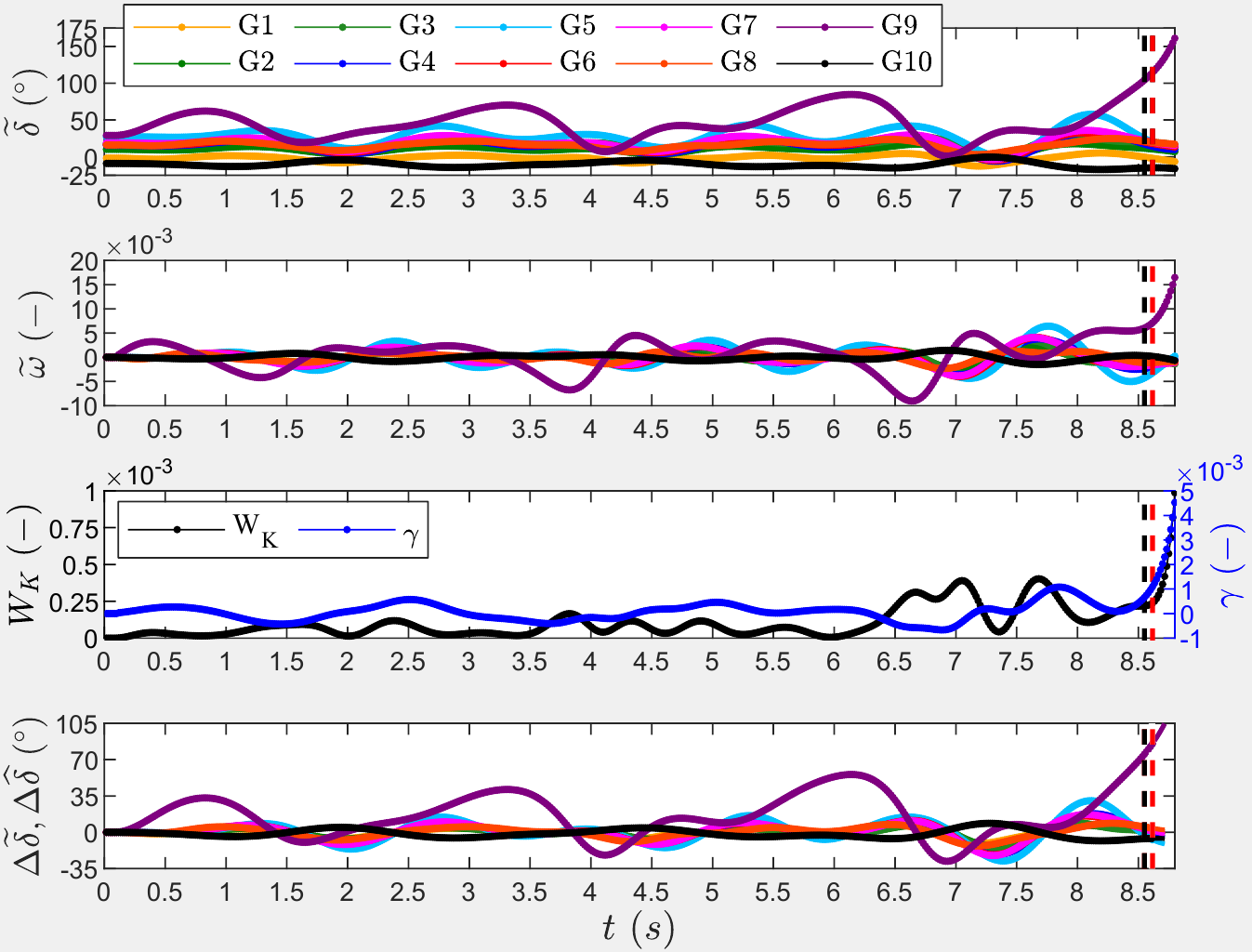}
\caption{Selected OOS indices following tripping of line 28--29}
\label{fig:flt2829}
\end{figure}
\begin{figure}[t]
\includegraphics[width=0.474\textwidth,left]{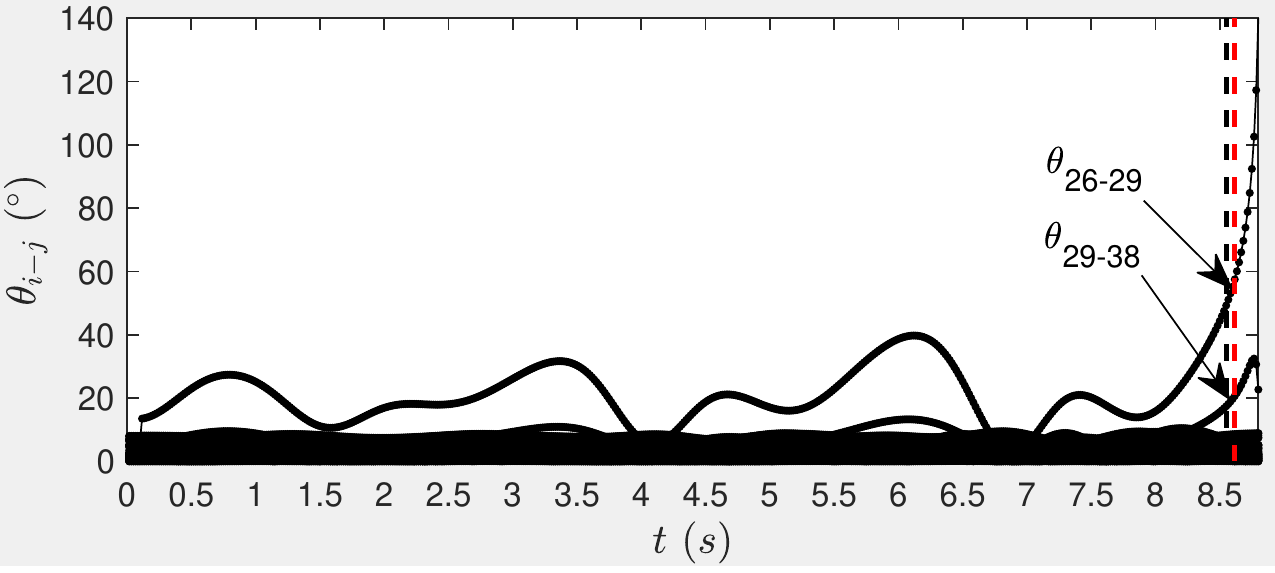}
\caption{Branch angle differences after tripping of line 28--29}
\label{fig:theta2829}
\end{figure}
\begin{figure}[t]
\includegraphics[width=0.474\textwidth,left]{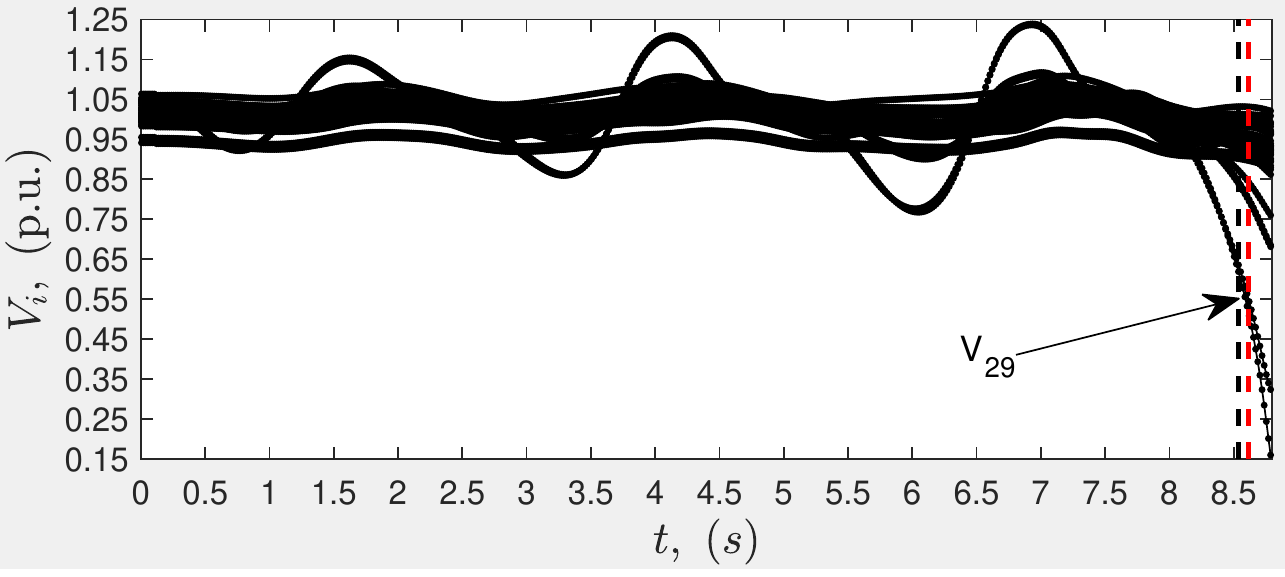}
\caption{Bus voltage magnitudes after tripping of line 28--29}
\label{fig:volt2829}
\end{figure}

\begin{figure}[t]
\centering
\includegraphics[width=0.4\textwidth]{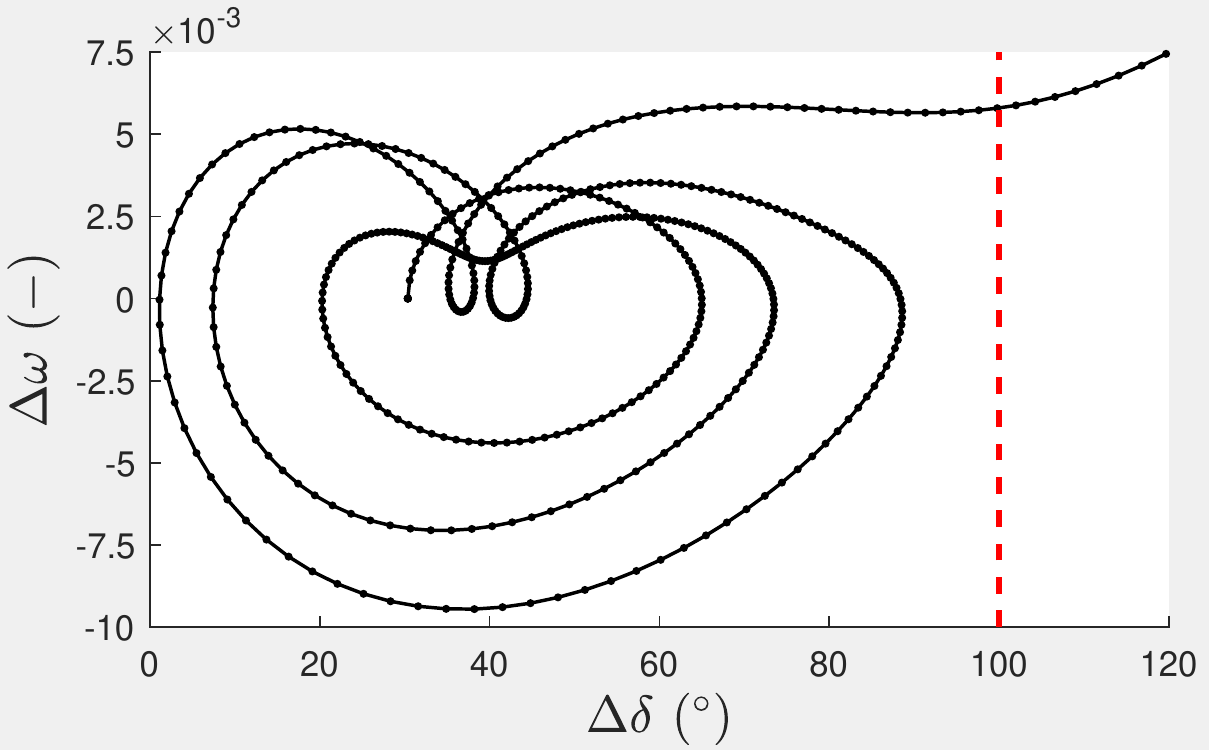}
\caption{Phase portrait of G9 following tripping of line 28--29}
\label{fig:phpt2829}
\end{figure}

For an example of a single generator instability and undamped oscillations, line 28--29 is switched off without any fault. As it can be seen from Figure \ref{fig:flt2829}, this results in undamped oscillations of generator 9 at bus 38, which eventually result in the out-of-step of that generator. The out-of-step is detected by \eqref{eqn:oosdct2} due to the growing system kinetic energy and the unstable phase portrait of generator 9 that can be seen in Figure \ref{fig:phpt2829} (the dashed red line in Figure \ref{fig:phpt2829} marks the chosen value of $\delta_{arm}^{\mathrm{COI}}$). The undamped power swings detector from Section \ref{sec:Wke} has counted four growing energy peaks for generators 7 and 9 and five growing energy peaks for generator 5. However, as discussed in Section \ref{sec:Wke}, the energy peaks counters $\kappa_i$ are suggested to be used for alarm purposes, as growing oscillations do not always result in an out-of-step. At the moment of OOS detection, $\theta^{\mathrm{max}}_{26-29}=57.5^{\circ}$, $\theta_{29-38}=20.5^{\circ}$ and the lowest bus voltage is 0.51 p.u. at bus 29, as it can be observed in Figures \ref{fig:theta2829}--\ref{fig:volt2829}. The predicted generator angles have low errors as in the previous two scenarios. Their pattern strongly suggests to separate generator 9 from the system, which is implemented by opening lines 25--26 and 17--27. Again, the cutset computed using \cite{Tyuryukanov.2018b} offers a chance to improve the islands' power balance. As obvious from Figure \ref{fig:theta2829}, local OOS protection would instead trip line 26--29, which would result in poorly balanced islands after splitting. Meanwhile, the formed islands \{9\} and \{1, 2, 3, 4, 5, 6, 7, 8, 10\} remain stable after splitting.

\section{Conclusions and Outlook}
\label{sec:conclus}
In this paper, we have proposed a data-driven controlled power system separation approach to mitigate the OOS conditions in a coordinated manner. It is designed to consistently answer the two most important questions related to controlled system separation: \emph{when to split} and \emph{where to split}. The "when to split" problem is systematically addressed by a set of instability indicators based on fundamental concepts (e.g., $W_K$, phase portraits, diverging generator angles). To decide where to split, CM are estimated, which is related to the well-known generator coherency problem. However, the proposed approach is more flexible, as its groups generators over much shorter time windows under the premise that the OOS has been detected and the generator trajectories are diverging. If the presence of undamped power swings has been determined (e.g., by our proposal in Section \ref{sec:Wke}), the known online coherency grouping algorithms could also be used, since long data windows become feasible in this case.

The key advantage of controlled splitting over traditional OOS protection is in its ability to act independently of the location of the oscillation center to improve the stability and power balance of the formed islands. Furthermore, the case studies on the IEEE 39-bus test system demonstrate that our scheme usually detects OOS significantly before it can be seen in the voltages, currents, and frequencies across the network. However, the requirement of separating certain network elements (e.g., the CM) makes controlled splitting inherently NP-hard to solve. For this reason, it cannot be seen as a complete substitute of local OOS protection. Local OOS relays should always serve as backup if the ongoing instability does not match any of the predefined CGGs. 

To relate the developed splitting scheme with some real instability scenarios, the recent ENTSO-E system splits can be examined. Their analysis \cite{ENTSOE.2021a,ENTSOE.2021b} shows a clear divergence in voltage angles and frequencies between the separated areas, which could be captured by the proposed OOS detectors. In case of the France-Spain separation, it could assist in issuing the tripping commands, as the generators of the Iberian peninsula form a distinct group with non-ambiguous boundaries. The liability issues due to the possible involvement of multiple utilities into controlled system separation could be mitigated by mutual special agreements. Such agreements already exist between various transmission system operators (TSO) within ENTSO-E for a number of SIPSs spanning several utilities. Possible concerns about the resilience of the proposed scheme (and of PMU-based wide-area protection schemes in general) against cyber attacks can also be largely alleviated. For example, it is already feasible to have multiple sources of time synchronization in a network of PMUs, which rely on different principles (e.g., GPS time signals and time synchronization via Ethernet or fiber optics by using the Precise Time Protocol \cite{Naglic.2020,Sheereen.2021}) for increased robustness.

Although the case studies in this paper involve standard IEEE benchmark power systems and conventional synchronous generators, it has been explained that the proposed splitting scheme could be used in power systems containing synchronous machines and grid-forming full-rated converters (FRC) as IBRs. Thus, our approach could support the ongoing replacement of conventional synchronous generators with IBRs in many power systems. Further investigating this topic could be a subject of future work.

\section*{Acknowledgment}
\label{sec:acknow}
The authors would like to specially acknowledge the contribution of Dr. Matija Naglič (currently at TenneT TSO B.V.) for his advice on hardware aspects of PMUs. Anonymous reviewers are highly acknowledged for their comments, which noticeably helped to improve the paper.

\bibliographystyle{IEEEtran}
\bibliography{references}

\vfill
\vfill
\clearpage
\end{document}